\documentclass[]{aa}
\usepackage{graphicx}

\begin{document}

\title{The $\sigma$\,Orionis substellar population.\thanks{
These observations were collected at the VLT of the 
European Southern Observatories} }

\subtitle{VLT/FORS spectroscopy and 2MASS photometry}

   \author{David Barrado y Navascu\'es
          \inst{1}
   \and
          V\'\i ctor J.\,S$.$ B\'ejar
          \inst{2}
   \and
          Reinhard Mundt
          \inst{3}
   \and
          Eduardo L$.$ Mart\'{\i}n
          \inst{4}
   \and
          Rafael Rebolo
          \inst{2,5}
   \and
          Mar\'{\i}a Rosa Zapatero Osorio
          \inst{1}
   \and
          Coryn A.\,L$.$ Bailer-Jones
          \inst{3}
   }

   \offprints{D. Barrado y Navascu\'es}

   \institute{Laboratorio de Astrof\'{\i}sica Espacial y F\'{\i}sica 
        Fundamental,
        INTA,  P.\,O$.$ Box 50727, E--28080 Madrid, Spain\\
        \email{barrado@laeff.esa.es}
   \and
        Instituto de Astrof\'\i{}sica de Canarias, E--38205 La Laguna, 
        Tenerife, Spain
   \and
        Max-Planck-Institut f\"ur Astronomie,
        K\"onigstuhl 17, D--69117 Heidelberg, Germany 
   \and
        Institute of Astronomy. University of Hawaii at Manoa. 
        2680 Woodlawn Drive, Honolulu, HI 96822, USA
   \and
        Consejo Superior de Investigaciones Cient\'{\i}ficas, CSIC, Spain
   }

   \date{Received; accepted }

   \abstract{VLT/FORS spectroscopy and 2MASS near-infrared photometry,
             together with previously known data, have been used to
             establish the membership and the properties of a sample
             of low-mass candidate members of the $\sigma$\,Orionis
             cluster with masses spanning from 1\,$M_{\odot}$ down to
             about 0.013\,$M_{\odot}$ (i.e., deuterium-burning mass
             limit). We have observed $K$-band infrared excess and
             remarkably intense H$\alpha$ emission in various cluster
             members, which, in addition to the previously detected
             forbidden emision lines and the presence of Li\,{\sc i}
             in absorption at 6708\,\AA, have allowed us to
             tentatively classify $\sigma$\,Orionis members as
             classical or weak-line T\,Tauri stars and substellar
             analogs. Variability of the H$\alpha$ line has been
             investigated and detected in some objects. Based on 
             the $K$-band infrared excesses and the
             intensity of H$\alpha$ emission, we estimate that the
             minimum disk
             frequency of the $\sigma$\,Orionis low-mass population is
             in the range 5--12\%.
%, which is expected for the age
%             ($\sim$3\,Myr) of the cluster.  
              }

  \titlerunning{The $\sigma$\,Orionis substellar population: H$\alpha$ and IR photometry}
%  \titlerunning{The $\sigma$\,Orionis substellar population}

  \authorrunning{Barrado y Navascu\'es et al.}

  \maketitle

\section{Introduction}

This paper is part of a series devoted to the study of the 
young, nearby open cluster associated with the $\sigma$ Orionis
multiple  star (O9.5 V spectral type). 
The clustering of B stars around $\sigma$ Orionis
 was noticed by Garrison (1967),
whereas  the cluster was listed by Lyng\aa{ }
(1981, 1987)  and  re-discovered by 
Wolk (1996) and Walter et al. (1997) using ROSAT data and
 spectroscopic and photometric follow-up.
This stellar association is characterized by its moderate closeness
(the Hipparcos distance is d=352$^{+166}_{-85}$ pc),
young age
($\tau$=4.2$^{+2.7}_{-1.5}$ Myr by Oliveira et al. 2002. See Zapatero
Osorio et al. 2002a for another estimate)
and low interstellar reddening 
(E($B-V$)=0.05, Lee 1968; Brown et al. 1994).

%-----------------------------------------------------------
    \begin{figure*}
    \centering
    \includegraphics[width=7.2cm]{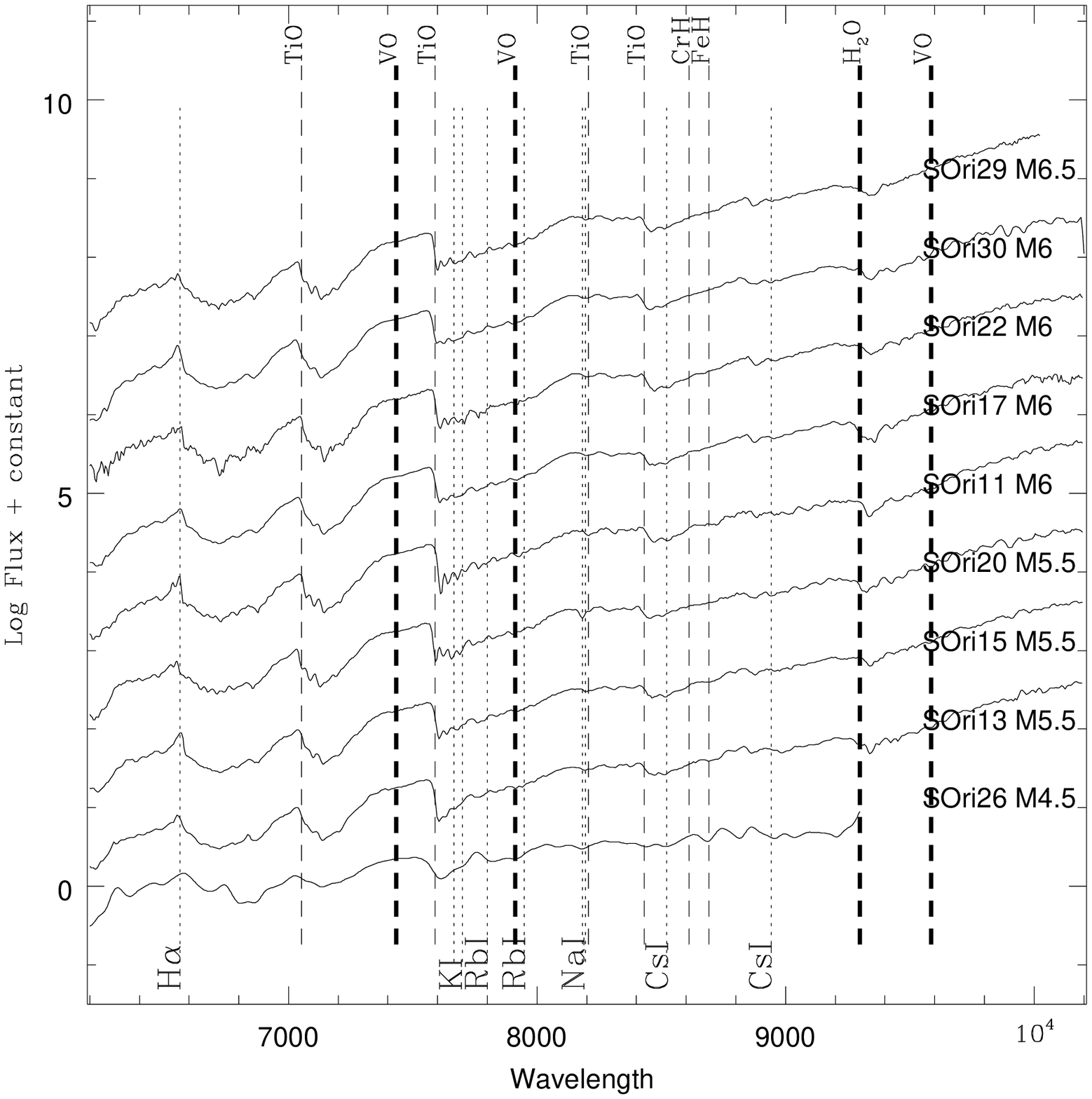}
    \includegraphics[width=7.2cm]{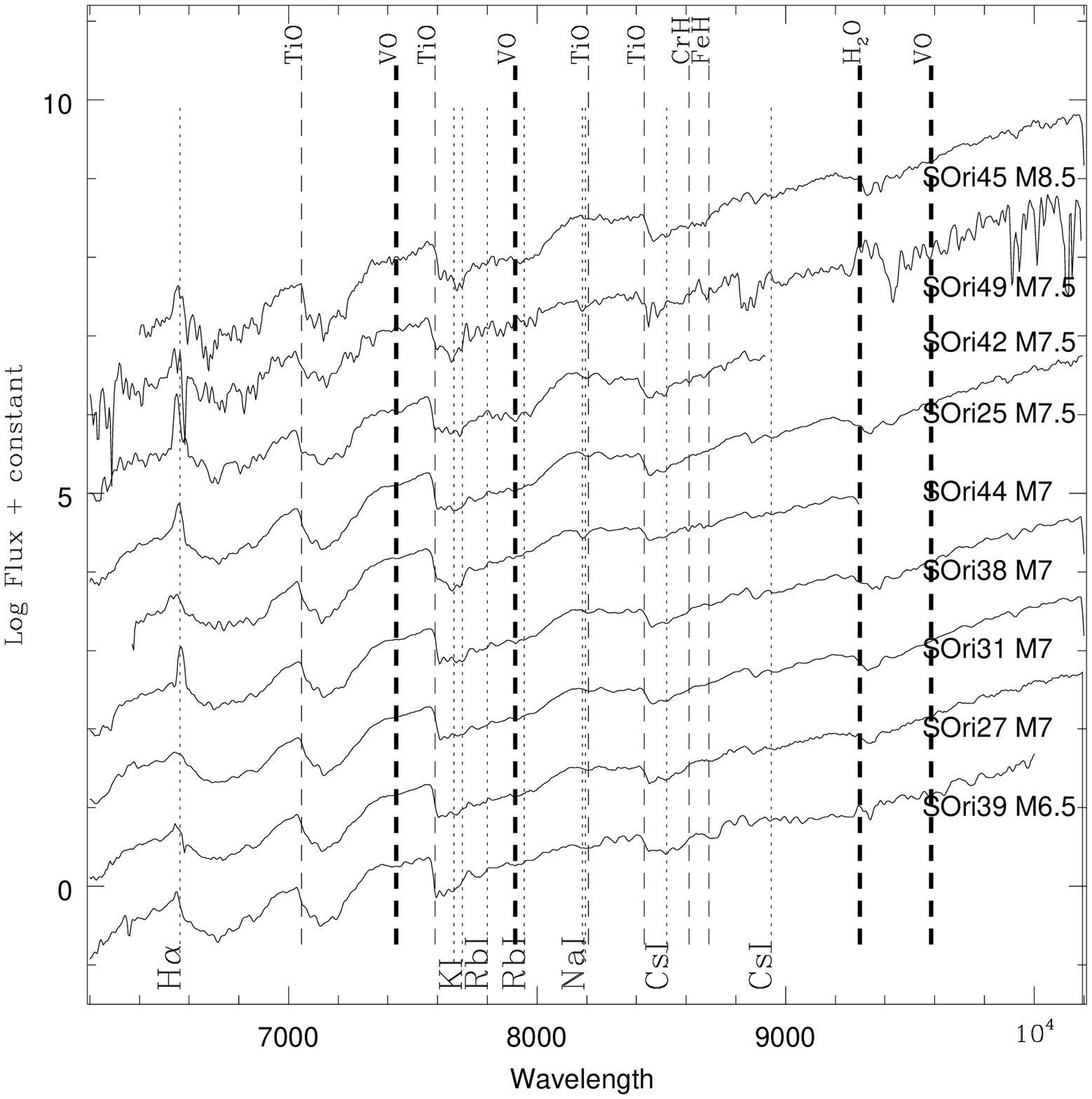}
    \includegraphics[width=7.2cm]{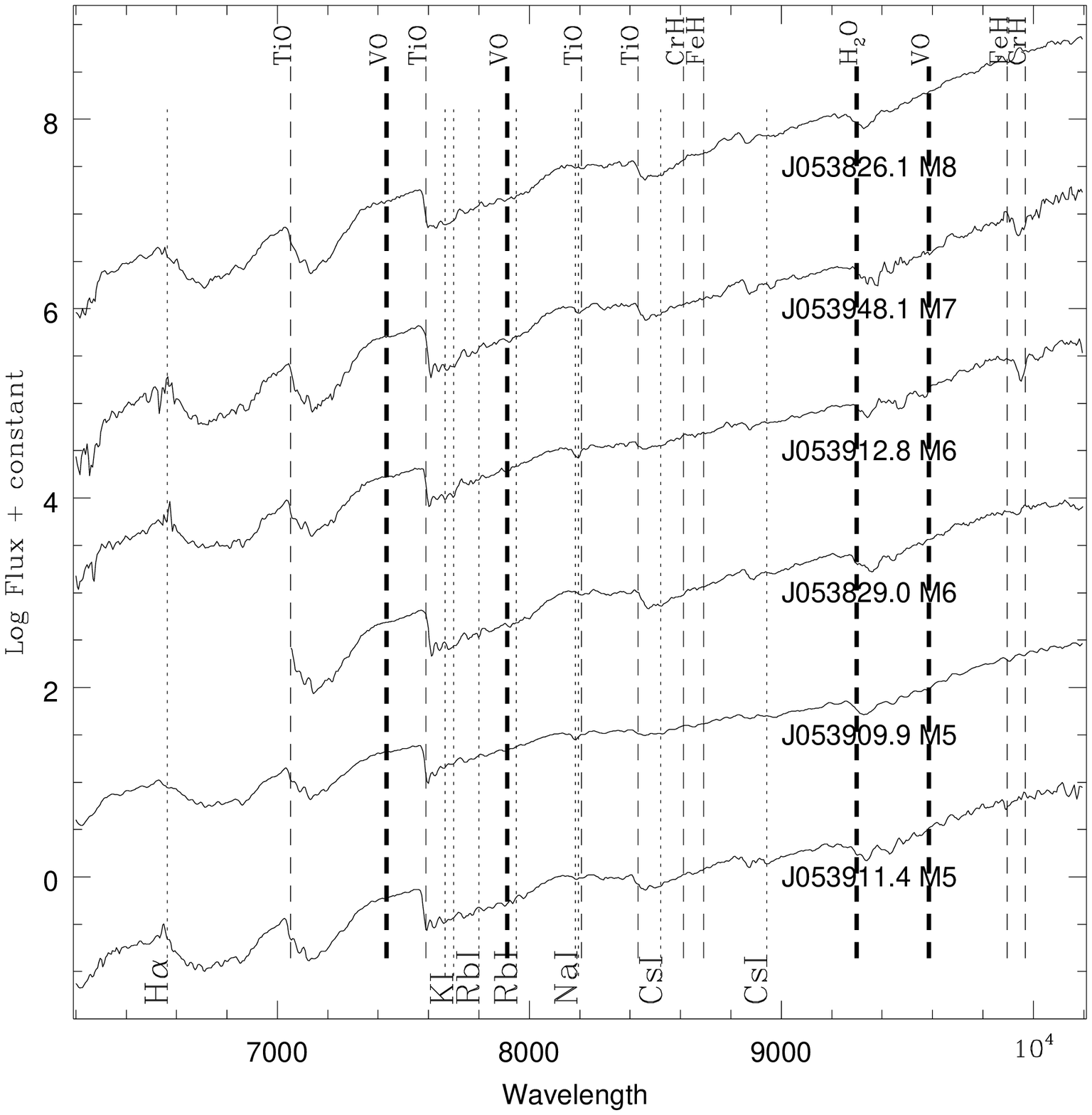}
 \caption{VLT/FORS  spectra of brown dwarf candidate members of 
          the $\sigma$\,Orionis cluster (ordered by decreasing 
          spectral type). Note the logarithmic scale in the 
          flux-axis.  }
    \end{figure*}
%______________________________________________________________      

So far, we have carried out a census of its population,
both stellar and substellar. 
Substellar objects can be subclassified as 
brown dwarfs (BDs) and isolated planetary--mass objects (IPMOs).
The first type is characterized by the lack of stable 
hydrogen burning
during any stage of their evolution, whereas the second group
is unable of any nuclear reaction of energetic significance 
at all, including the 
deuterium  burning. For solar metallicity, 
the borderlines have been computed as
$\sim$0.075 M$_\odot$ and $\sim$0.013 M$_\odot$, respectively
(Kumar 1963; D'Antona \& Mazzitelli
1994, 1997; Saumon et al. 1996; 
Burrows et al. 1997; Chabrier et al. 2000).
The initial searches of substellar components 
in the $\sigma$ Orionis cluster 
were presented in B\'ejar et  al. (1999) and
Zapatero Osorio et al. (1999).
The discovery of  IPMOs in this association,
by means of photometric searches and the spectroscopic 
confirmation of the cool nature of three of them, 
was presented in Zapatero Osorio et al. (2000), whereas 
their substellar  nature
was established via optical and infrared spectroscopy
in  Mart\'{\i}n et al. (2001) and Barrado y Navascu\'es et al. (2001).
In this last paper, we presented the 
first detection of  H$\alpha$ emission in IPMOs.
The substellar Initial Mass Function (IMF) of the cluster 
was derived by  B\'ejar et al. (2001) and 
binarity was investigated in  Mart\'{\i}n et al. (2001).
In Zapatero Osorio et al. (2002a),
we established that the most likely  age of the cluster
is 3 Myr,  and it is effectively bracketed between 1 and 8 Myr, 
based on the 
evolution of their eponymous star and the lithium photospheric content
of its low mass members. Additionally, we 
detected  H$\alpha$ emission in most of the very low mass stars
and BDs, as well as several forbidden lines 
([OI]6300\AA, [OI]6364\AA, [NII]6548\AA, 
[NII]6583\AA, [SII]6716\AA, [SII]6731\AA) in about a third
of objects belonging to this sample,  indications of accretion
and/or mass ejection.
Finally, in Zapatero Osorio et al. (2002b) and Barrado y Navascu\'es
et al. (2002a), we investigated  the 
presence of very strong   H$\alpha$ emission of unknown origin
in  two members with masses around the planetary-domain  limit
 (S\,Ori\,55 and S\,Ori\,71).

This paper complements our previous works 
on H$\alpha$ emission, focussing on 
 objects with masses around
the substellar limits at about 0.075 M$_\odot$ and those with masses above the
 deuterium border at about 0.013 M$_\odot$.  
In the present paper, we derive the spectral types,
study the infrared properties and 
try to disentangle the origin  of the
detected H$\alpha$ emission line 
in a sample of brown dwarfs belonging to the 
$\sigma$ Orionis cluster, discussing the possible presence of
 accretion disks in a  fraction of them.

%-----------------------------------------------------------
    \begin{figure}
    \centering
    \includegraphics[width=8.8cm]{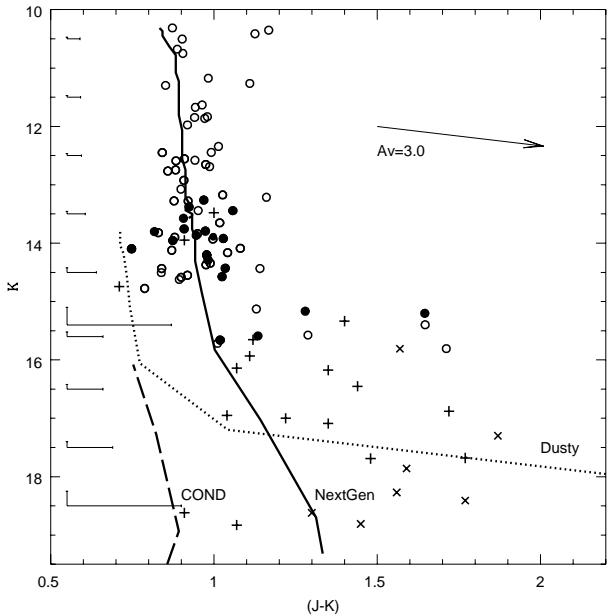}
 \caption{Infrared photometry for low-mass stars, brown dwarfs and
 planetary-mass objects of the $\sigma$\,Orionis cluster. Circles
 correspond to 2MASS data (solid symbols represent data with VLT
 spectroscopy). Data from B\'ejar et al$.$ (2001) and Mart\'{\i}n et
 al$.$ (2001) are plotted with crosses and plus symbols,
 respectively. Models by Chabrier et al$.$ (2000) and Baraffe et al$.$
 (1998, 2002) are included in the figure with a thick solid line
 (dust-free), dotted line (dusty) and long-dashed line (condensed).  }
 \end{figure}
%______________________________________________________________      

%-----------------------------------------------------------
    \begin{figure}
    \centering
    \includegraphics[width=8.8cm]{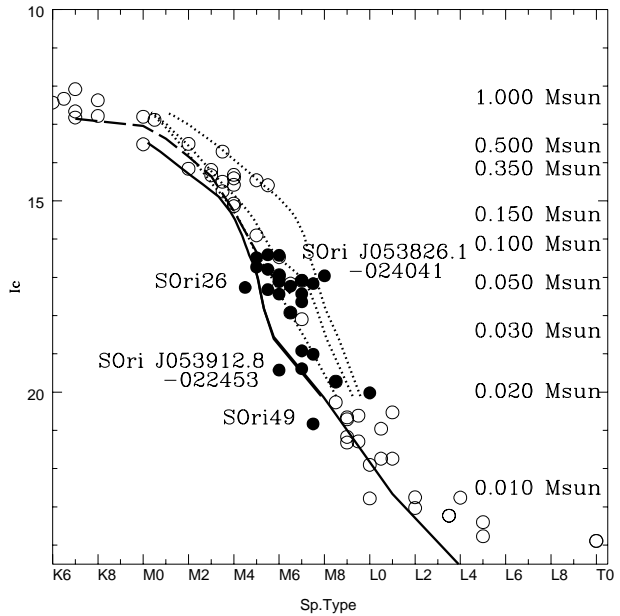}
 \caption{$I_c$ magnitude against spectral type for the
   $\sigma$\,Orionis cluster.  Solid circles represent data from this
   study, whereas open circles correspond to data from Barrado y
   Navascu\'es et al$.$ (2001), Mart\'{\i}n et al$.$ (2001), B\'ejar
   et al$.$ (2001) and Zapatero Osorio et al$.$ (2002a).  The location
   of the substellar borderline at the cluster age and distance is at
   around $I_c$\,=\,16\,mag.  The lines represent several
   3\,Myr-isochrones from Baraffe et al$.$ (1998), which were obtained
   for different temperature scales (high gravity by Basri et al$.$
   2000---solid line; different gravities by Luhman 1999---dotted
   lines, and Bessell 1991---dashed line).  Masses are also indicated
   with the labels in the right hand-side of the diagram.} 
    \end{figure}
%______________________________________________________________

\section{Observations}

\subsection{Red optical spectroscopy}
Spectra were collected at the Very Large Telescope Unit \#1 on the
Paranal Observatory of the European Southern Observatory during 2000,
December 23--27. We used the FORS1 spectrograph and the multi-slit
capability. FORS1 has a 0.2\arcsec/pixel scale in the standard
resolution, and yields a field of view of
6.8\arcmin$\times$6.8\arcmin. We used the 150I grism and the
order-blocking filter OG590. With a slit width of 1.4\arcsec, the
resolution is R$\sim$250 for the spectral coverage of our data. All of
our targets are $\sigma$\,Orionis brown dwarf candidates selected from
the surveys of B\'ejar et al$.$ (1999, 2001), except for a few, which
belong to the survey described in B\'ejar et al$.$ (2003, in
preparation). We optimized the multi-slit masks to include as many
faint, red candidate members as possible. Individual exposure times
were 2400 seconds. In the particular case of
S\,Ori\,13, we only collected data during 1290 seconds.  For
S\,Ori\,11, 15, 20, 27, 38, and 39, we obtained a sequence of up to 7
consecutive individual spectra.  Table~1 lists magnitudes, optical and
infrared colors, as well as the derived spectral types (see section
3.1) and other useful information of our targets.

The data were reduced using standard procedures within the
IRAF\footnote{IRAF is distributed by National Optical Astronomy
Observatories, which is operated by the Association of Universities
for Research in Astronomy, Inc., under contract to the National
Science Foundation, USA} environment.  Additional information on data
reduction can be found in Barrado y Navascu\'es et al$.$ (2001), where
we presented the VLT spectra of many $\sigma$\,Orionis planetary-mass
objects.  Whenever several exposures were available, the images were
added together before the spectrum extraction, producing an average
two dimensional spectrum image.  Then, in both the average and
individual frames, the spectra were extracted using the ``apall''
algorithm within IRAF.  We removed the sky emission lines and
background contributions by fitting the sky during the extraction
procedure.  The wavelength calibration was performed using
He\,Ar\,Hg\,Cd comparison arcs taken with the same instrumental
configuration.  Then, data were flux calibrated using several
spectrophotometric standard stars.  The final signal-to-noise ratios
(S/N), as measured in the 7300--7600\,\AA{ } region, are in the range
150--20, corresponding to the brightest objects ($I$=16.4--17.5) or
targets with multiple exposures and to the faintest objects
($I$=19.01--20.83) with only one exposure (S\,Ori\,42, 45, and~49).
Figure~1 displays the spectra of our $\sigma$\,Orionis brown dwarf
candidate members. In addition to the $\sigma$\,Orionis targets and
for comparison purposes, we also observed several nearby field objects
of M and L spectral types drawn from the lists of Kirkpatrick et al$.$
(1999) and Mart\'{\i}n et al$.$ (1999).
%-----------------------------------------------------------
    \begin{figure}
    \centering
    \includegraphics[width=8.8cm]{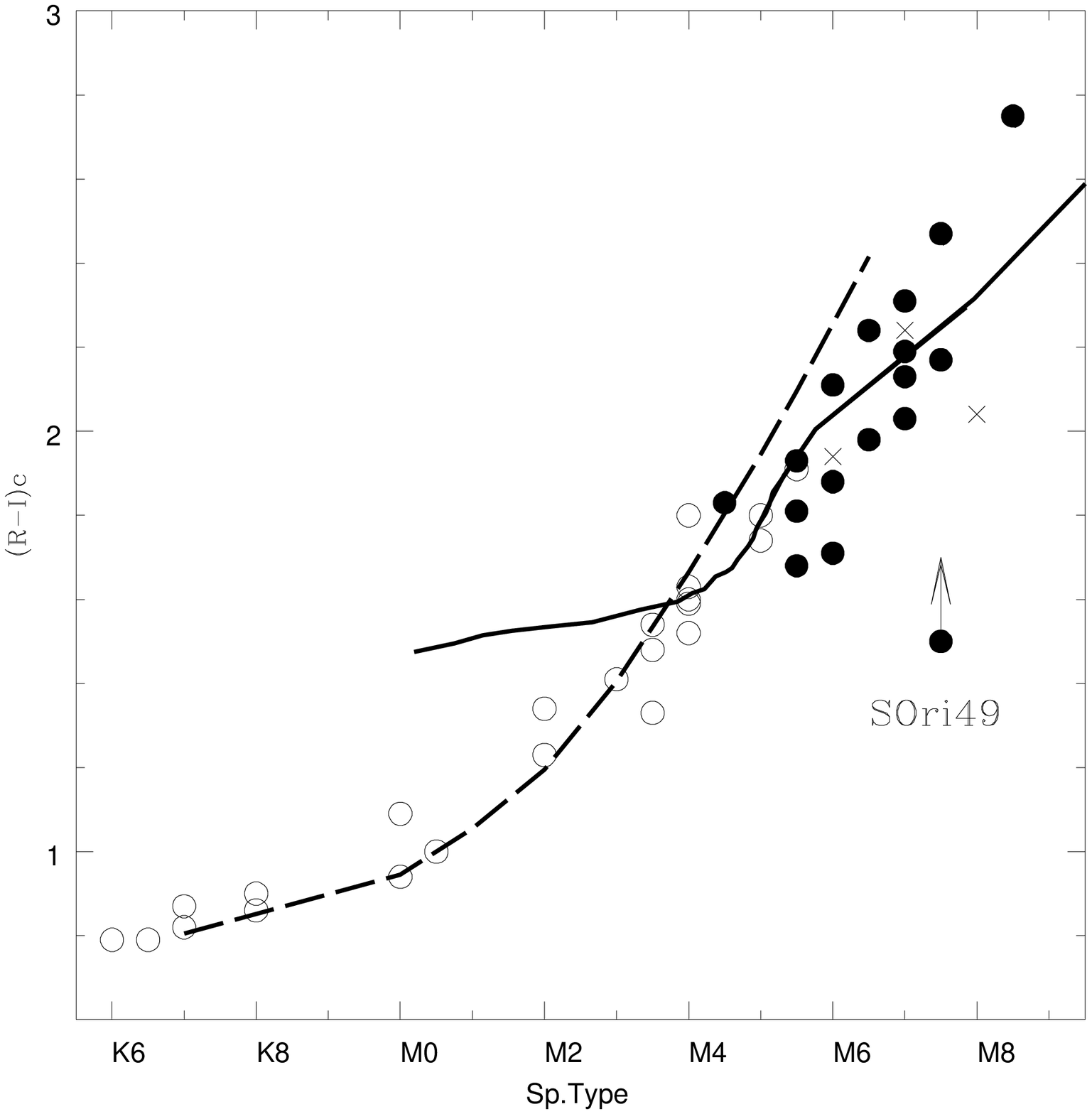}
    \includegraphics[width=8.8cm]{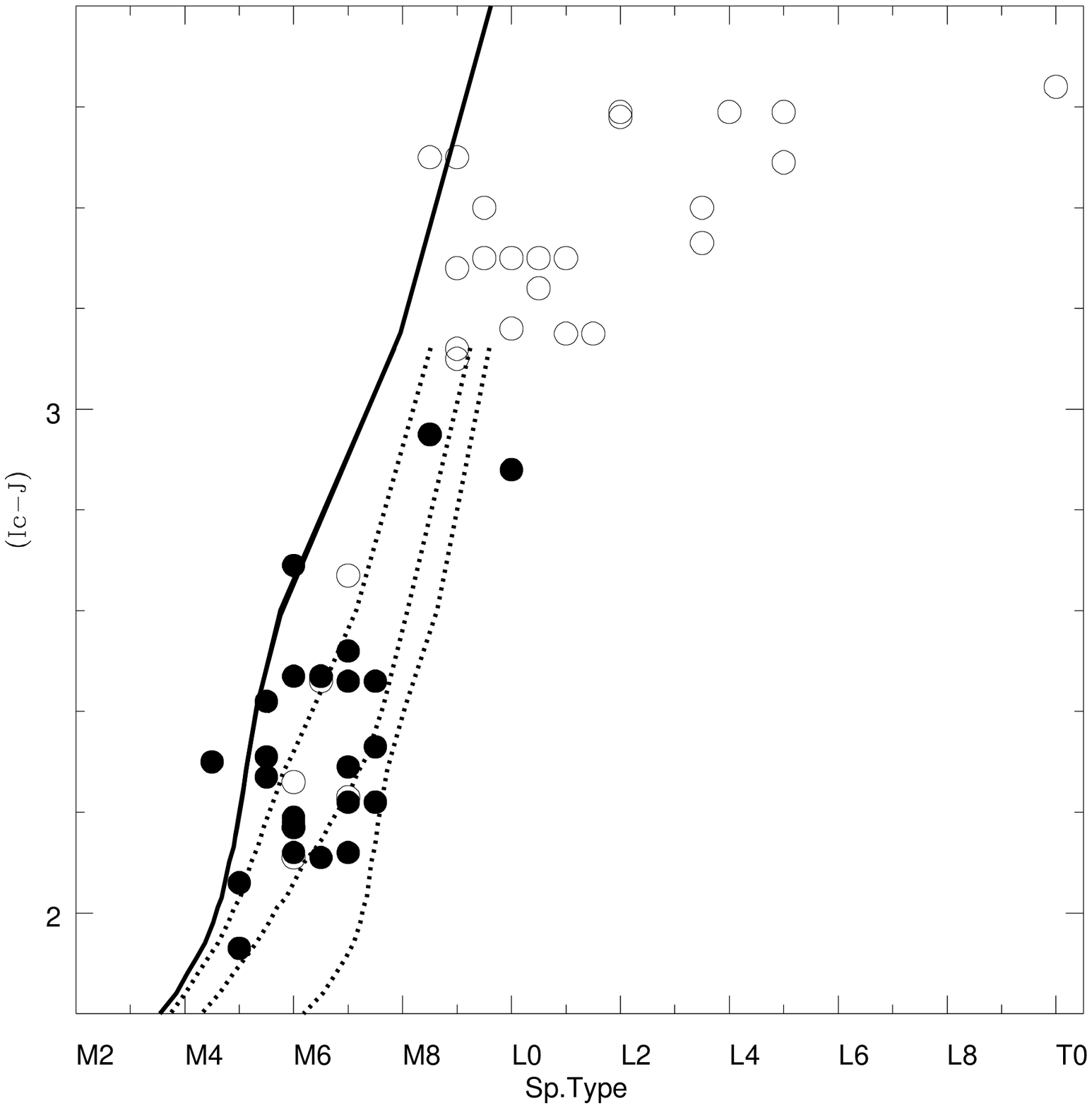}
 \caption{ Spectral type against $(R-I)_c$ color (upper panel) and
   $(I_c-J)$ color (bottom panel).  Solid circles represent data from
   this paper, whereas open circles correspond to data from B\'ejar et
   al$.$ (2001), Mart\'{\i}n et al$.$ (2001), Barrado y Navascu\'es
   et al$.$ (2001) and Zapatero Osorio et al. (2002a).
  The lines represent several 3\,Myr isochrones
   from Baraffe et al$.$ (1998), which were obtained for different
   temperature scales (Basri et al$.$ 2000---solid line, and Bessell
   1991---dashed line, in the upper panel; Basri et al$.$ 2000---solid
   lines, and Luhman 1999---dotted lines in the bottom panel).  }
 \end{figure}
%______________________________________________________________      

\subsection{Near-infrared photometry}
We also searched the 2MASS point source catalog (Skrutskie et al$.$ 1997),
 second incremental release, to identify the near-infrared
counterparts of $\sigma$\,Orionis low-mass stars and brown dwarfs from
our lists of optical sources. We used a searching radius of
5\arcsec. The results are listed in Table~2.  The offsets between
the optical and infrared coordinates are given in column \#7, and the
coordinates according to the 2MASS catalogue, more accurate than
previously published positions, are provided in column \#8.  The 2MASS
coordinates are generally within the error bars of the ``optical''
astrometry.  Figure~2 illustrates the optical-near-infrared
color-magnitude diagram.  2MASS data are represented with circles
(solid circles for objects with VLT spectroscopy presented in this
paper); other data taken from B\'ejar et al$.$ (2001) and
Mart\'{\i}n et al$.$ (2001) are plotted with the plus and cross
symbols, respectively.
Error bars (only $-$$\Delta$$K$ and $+$$\Delta$$(J-K)$ are depicted) are
displayed. 
  Three isochrones, corresponding to an age of
3\,Myr, and representing non-dusty, dusty and condensed models from
the Lyon group (Baraffe et al$.$ 1998; Chabrier et al$.$ 2000) are
also displayed in the figure. We remark the peculiar behaviour of the
$\sigma$\,Orionis photometric sequence at around $K$\,=\,18\,mag,
where the $J-K$ color turns to blue values. This effect is supposed
to be explained by the settlement of atmospheric dust particles at the
bottom of the object's photosphere (see Mart\'{\i}n et al$.$ 2001, and
references therein).

 One object, namely S\,Ori\,47, has $JHK$ data
coming from United Kingdom infrared telescope (UKIRT). These data were
taken under photometric conditions in February 2000.

We have used the following values of cluster distance and interstellar
reddening to produce various figures of this paper:
d=352$^{+166}_{-85}$ pc,
E($B-V$)=0.05 (Lee 1968; Brown et al. 1994),
E($R-I$)$_c$=0.035,
E($I_c-J$)=0.049,
E($I_c-K$)=0.076,
E($J-K$)=0.027,
$A_V$=0.156, 
$A_I$=0.093, 
$A_J$=0.044, and 
$A_K$=0.017.
Many of these values have been derived from the interstellar
extinction law and transformation equations between filters and
photometric systems published by Rieke \& Lebofsky (1985) and Taylor
(1986).

%-----------------------------------------------------------
    \begin{figure*}
    \centering
    \includegraphics[width=7.2cm]{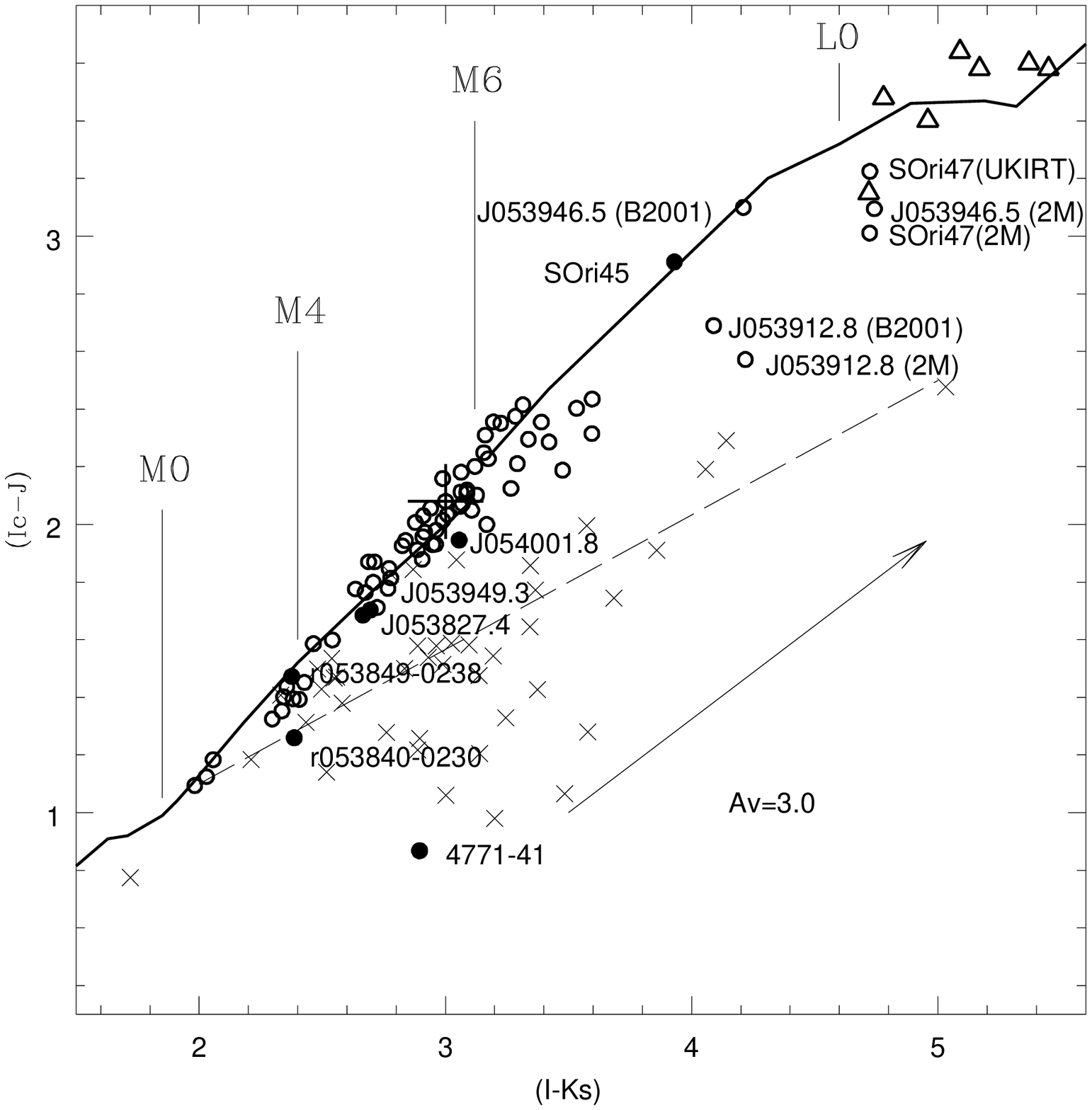}
    \includegraphics[width=7.2cm]{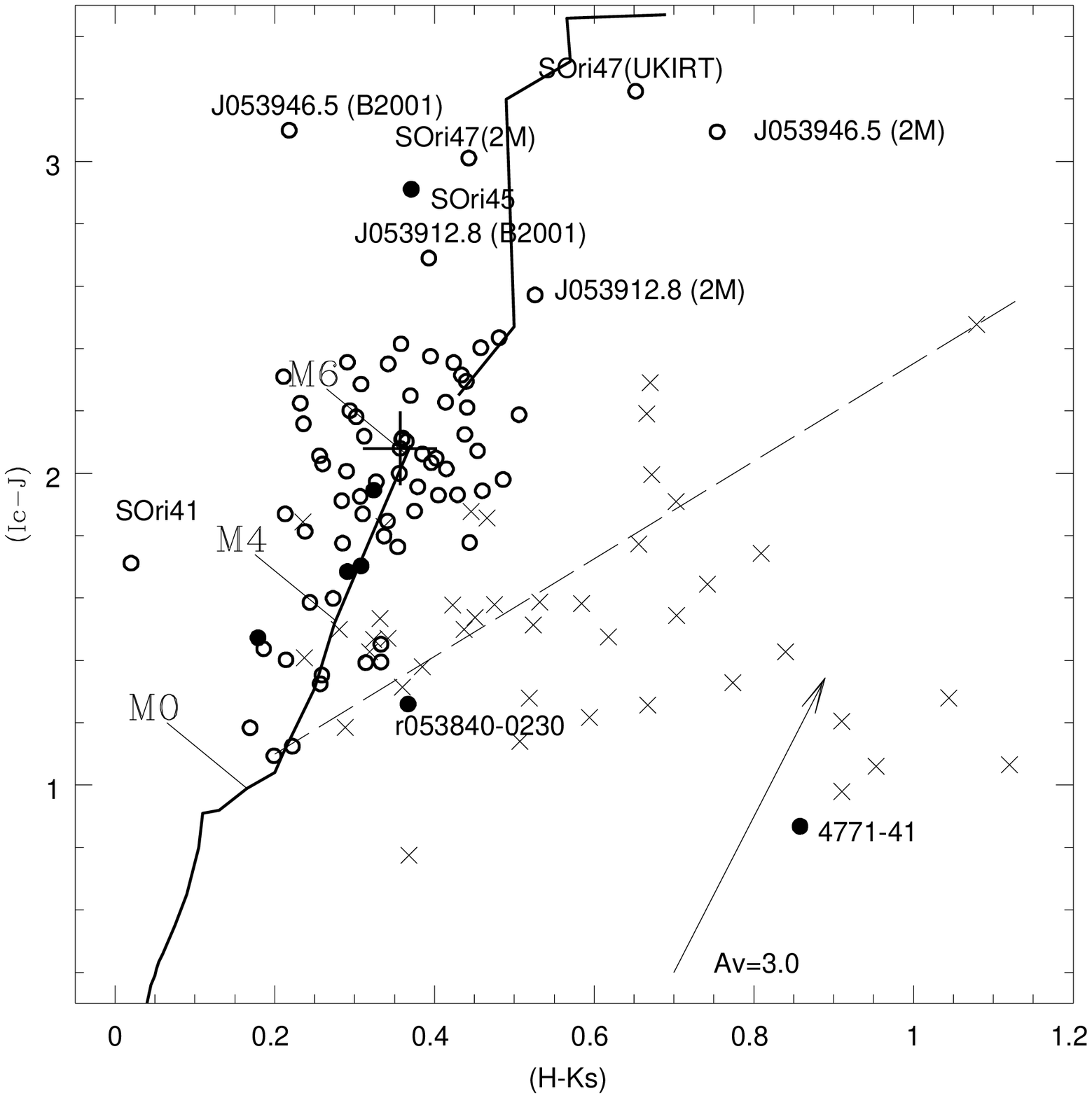}
    \includegraphics[width=7.2cm]{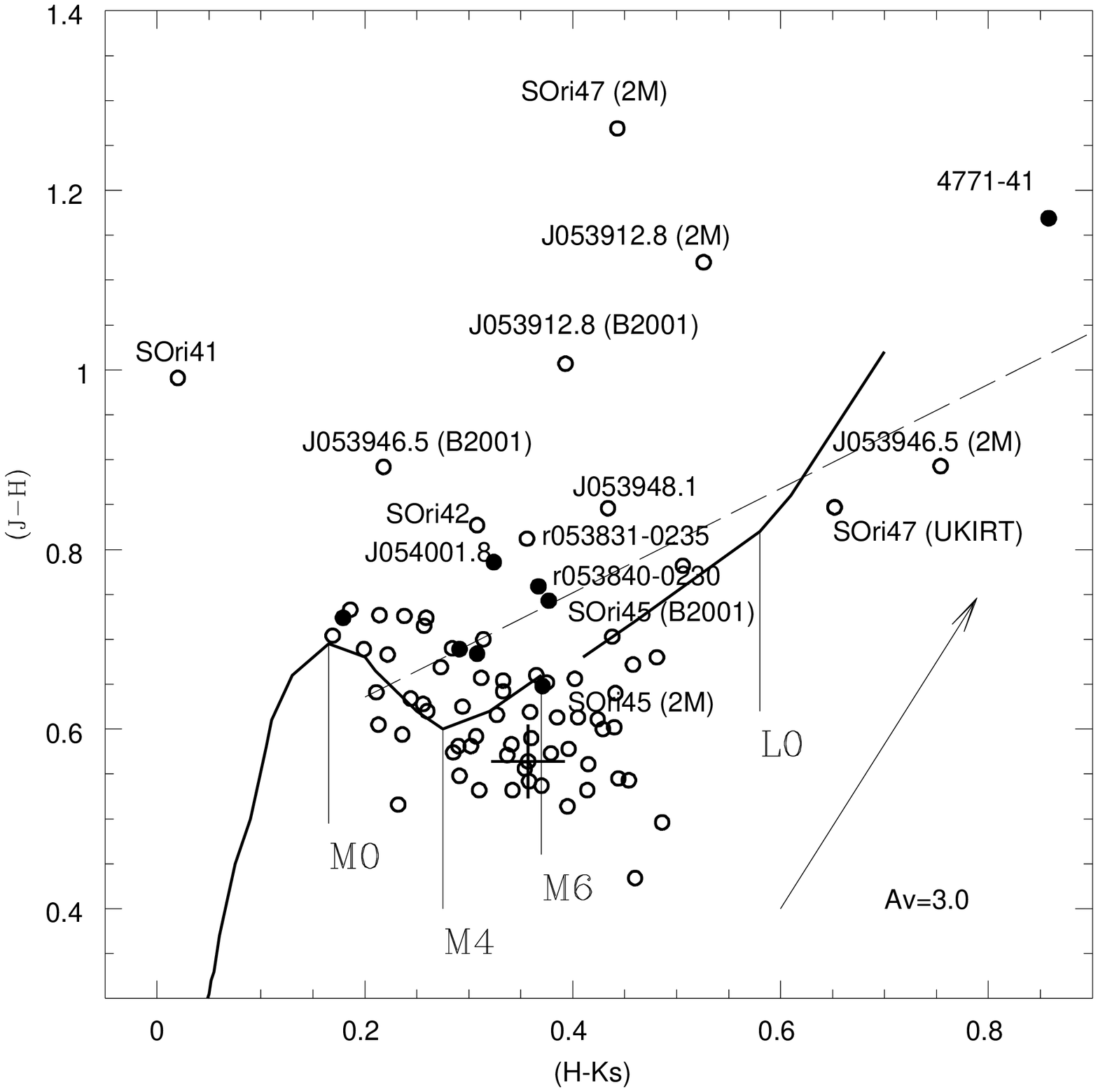}
 \caption{Optical-Infrared color-color diagrams based on 2MASS data for the
$\sigma$ Orionis population (open circles). 
Solid circles represent objects with 
forbidden emision lines. 
Open triangles correspond to isolate planetary mass objects.
 Crosses indicate the position of classical TTauri stars belonging to Orion
stellar population (Herbig \& Bell 1988).
The position of S\,Ori\,12 is indicated with a large plus symbol.
The thick-solid and dashed lines correspond to the locii of the main
 sequence stars (from Bessell \& Brett 1988; Kirkpatrick et al$.$
 2000; Leggett et al. 2001) 
and CTT stars (Meyer et al$.$ 1997 and this paper), respectively.  }
 \end{figure*}
%______________________________________________________________

\section{Analysis and discussion}

\subsection{Spectral types}
In Figure~1 we have included the identification of the major atomic
and molecular spectroscopic features characteristic of late-M and L
spectral classes, like K\,{\sc i} $\lambda$7665 and
$\lambda$7699\,\AA, Na\,{\sc i} $\lambda$8183 and $\lambda$8195\,\AA,
Rb\,{\sc i} $\lambda$7800 and $\lambda$7948\,\AA, Cs\,{\sc i}
$\lambda$8521 and $\lambda$9843\,\AA, and absorption bands of TiO,
FeH, CrH, VO and H$_2$O. The differences in the slope of the
pseudo-continuum and the change in the strength and width of the VO
and TiO bands can be appreciated in the VLT spectra of Figure~1.
Kirkpatrick et al$.$ (1999) and Mart\'{\i}n et al$.$ (1999) provide a
complete summary of the properties of very cool optical spectra of
field objects.  We can classify our $\sigma$\,Orionis candidates based
on their spectroscopic criteria. Flux ratios and spectral indices that
account for the pseudo-continuum slope have been measured over the
observed VLT data, and we have compared these measurements to those of
well-known spectroscopic standard stars to derive spectral types. Our
final classification, given in Table~1, ranges from M5 down to
M8.5. The uncertainty is estimated at half a subclass.  Seven out of
the total of 25 VLT objects have also been observed to a higher
spectroscopic resolution by B\'ejar et al$.$ (1999). Their typing,
which fully agrees with our classification considering error bars, is
included in Table~1.

%______________________________________________________________      

    \begin{figure*}
    \centering
    \includegraphics[width=14cm, angle=-90.0]{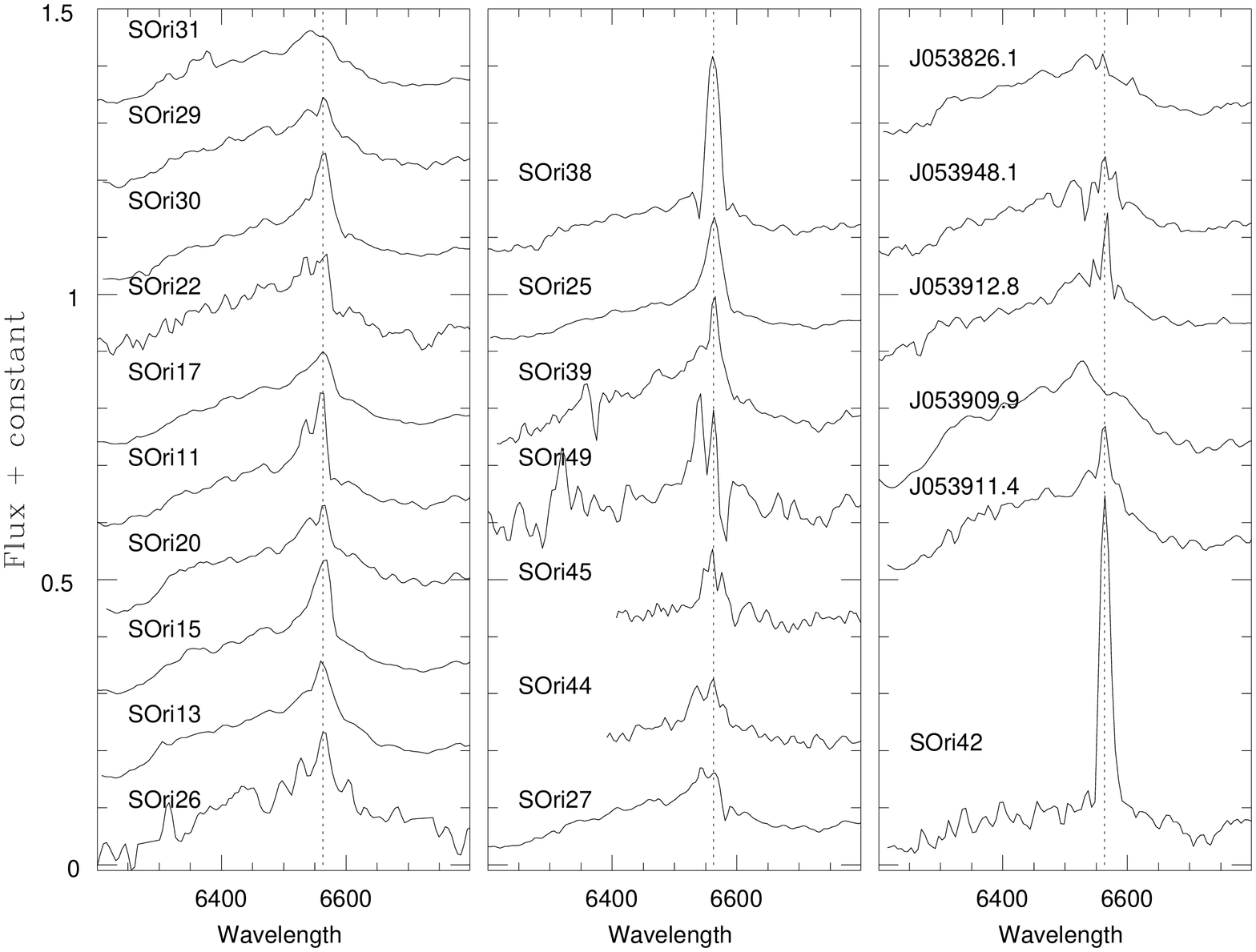}
 \caption{Details around the H$\alpha$ line.  When several spectra
  were available, we computed the average, which is the one shown in
  the figure.}
    \end{figure*}
%______________________________________________________________      

\subsection{Photometric and spectroscopic membership}
Given the very low resolution of our spectra, we cannot obtain
accurate radial velocities or see the atomic absorption features due
to lithium at 6708\,\AA{} and sodium at 8195\,\AA, which are
age-indicators.  As discussed in Zapatero Osorio et al$.$ (2002a), all
stars and brown dwarfs of the $\sigma$\,Orionis cluster preserve
lithium in their atmospheres. Instead, we have combined spectral types
and optical and near-infrared photometry to study the cluster
membership of our candidates. The H$\alpha$ emission line, which is
also a sign--post of youth, is discussed in section 3.4.

Figure~3 displays $I_c$ magnitudes against spectral types for the
$\sigma$\,Orionis cluster, and Figures~4a and 4b illustrate the
relation between the $(R-I)_c$ and $(I_c-J)$ colors and the spectral
classification, respectively. Data from B\'ejar et al$.$ (1999, 2001),
Barrado y Navascu\'es et al$.$ (2001), Mart\'{\i}n et al$.$ (2001) and
Zapatero Osorio et al$.$ (2002a) are plotted with open circles,
whereas the new data of this paper are shown as filled
circles. Overplotted is the 3\,Myr-isochrone from Baraffe et al$.$
(1998), which provides magnitudes and colors in the filters of
interest. This evolutionary isochrone has been converted into spectral
type by using the temperature scales of Bessell (1991), Basri et al$.$
(2000), and Luhman (1999). Based on the model, the mass range of the
VLT objects spans from the substellar frontier (0.075\,$M_\odot$) down
to roughly the planet--brown dwarf boundary (0.013\,$M_\odot$).

These diagrams show a clear continuous sequence of $\sigma$\,Orionis
stellar and substellar objects between spectral types K6 and
T0. Actually, the cluster spectroscopic sequence extends into the
T-class with the finding of S\,Ori\,70, which has been tentatively
classified as T6 (Zapatero Osorio et al$.$ 2002c). This object is not
included in Figures~3 and 4b for clarity. Several  objects stand
out of the cluster sequence, either because they are too faint or too
bright for their spectral type, or because their colors do not match
the measured spectral type. The objects that are below the sequence
are S\,Ori\,49, S\,Ori\,26 and S\,Ori\,J053912.8--022453. A possible
interpretation for their location in Figure~3 is that they are
spurious members of $\sigma$\,Orionis, cool objects in the
neighborhood of the cluster. However, we cannot totally rule out their
membership in $\sigma$\,Orionis because these objects may exhibit
strong differential reddening, or optical veiling due to accretion
from disks (like PC\,0025$+$0447, Mart\'{\i}n et al$.$ 1999).  This
is most likely the case of S\,Ori\,J053912.8--022453, since it has
significant near-infrared excesses (about A$_V$$\sim$3\,mag,
corresponding to A$_I$$\sim$1.8, see Figure~5).  Its de-reddened
location in Figure~3 agrees with S\,Ori\,J053912.8--022453 being a
true member of the $\sigma$\,Orionis cluster. On the other hand,
S\,Ori\,49 and S\,Ori\,26 lack strong H$\alpha$ emission, a feature
that is characteristic of young, cool objects (section 3.4). This may
indicate that they are non-members of the cluster. This result agrees
with B\'ejar et al$.$ (2001), who discarded S\,Ori\,49's cluster
membership based on the object's near-infrared photometry. Thus, only
2 objects are found to be contaminants among the 25 sources of our
sample, i.e., a pollution rate less than  10\%.

S\,Ori\,J053826.1--024041 (M8) and S\,Ori\,71 (L0) appear quite bright
for their spectral types in Figure~3, which may suggest, apart from
the fact that they might be younger than the cluster, that they are
binary cluster members (with equal mass components).  This situation
is similar to that of S\,Ori\,47 (L1, Zapatero Osorio et al. 1999), a
$\sigma$\,Orionis member close to the borderline between brown dwarfs
and planetary-mass objects (Barrado y Navascu\'es et al$.$ 2001). The
estimated mass of S\,Ori\,71 is also close to the planetary
boundary. If the double nature of S\,Ori\,71 and S\,Ori\,47 is finally
proved, they would become the first planetary-mass binary
systems. S\,Ori\,71 shows an incredibly intense H$\alpha$ emission. A
comprenhensive analysis of this object can be found in Barrado y
Navascu\'es et al$.$ (2002a).

%-----------------------------------------------------------
    \begin{figure}
    \centering
    \includegraphics[width=8.8cm]{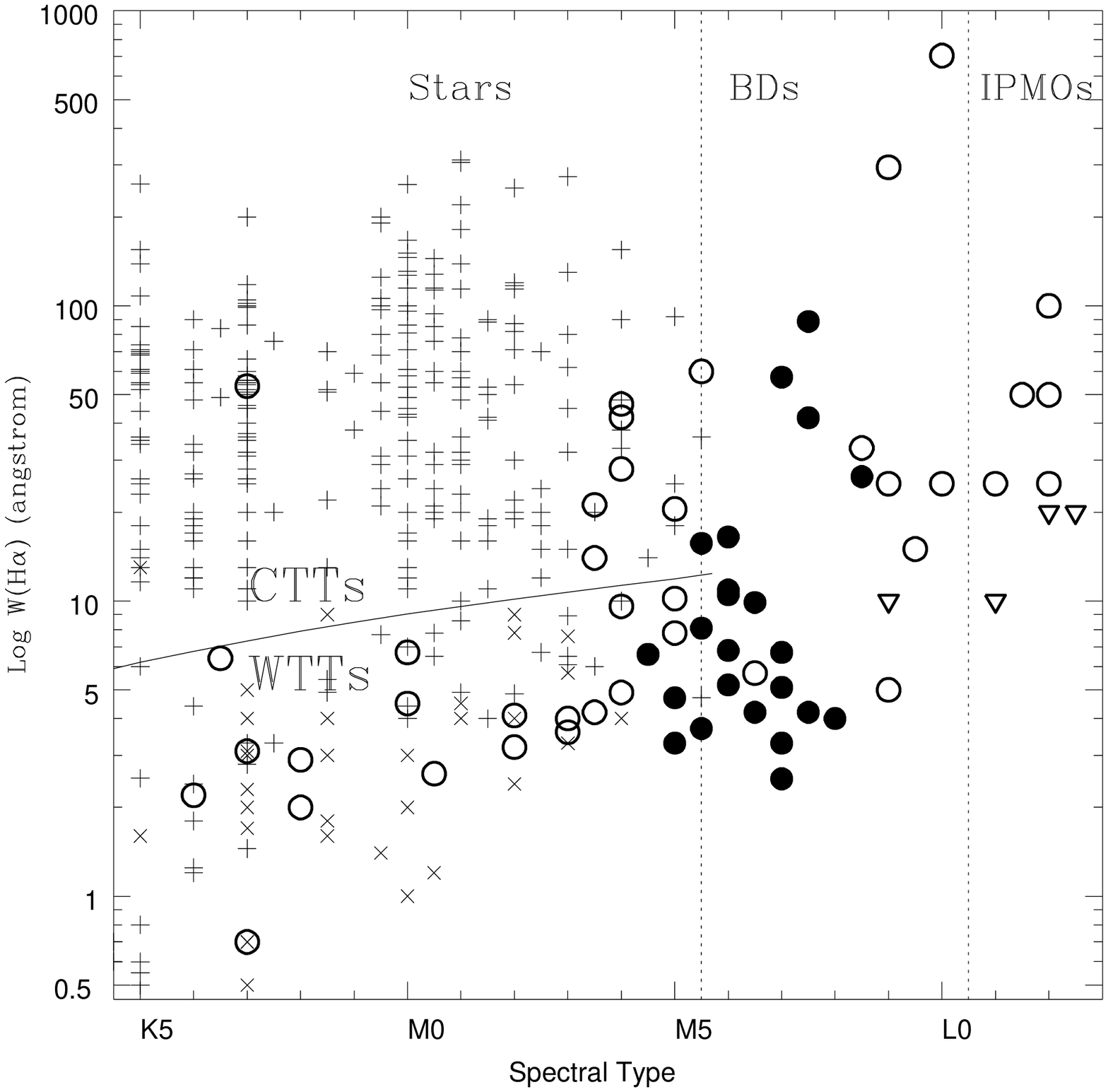}
 \caption{H$\alpha$ pseudo-equivalent width versus spectral type.
 Detections and upper limits in the $\sigma$\,Orionis cluster appear
 as circles and triangles, respectively.  Solid circles correspond to
 the data of this paper.  Open symbols represent data from B\'ejar et
 al$.$ (1999), Barrado y Navascu\'es et al$.$ (2001) and Zapatero
 Osorio et al$.$ (2002a). Crosses correspond to data of pre-main
 sequence stars of Orion (Herbig \& Bell 1988).  The solid line
 separates CTT from WTT stars.  } 
    \end{figure}
%______________________________________________________________      

%-----------------------------------------------------------
    \begin{figure}
    \centering
    \includegraphics[width=8.8cm]{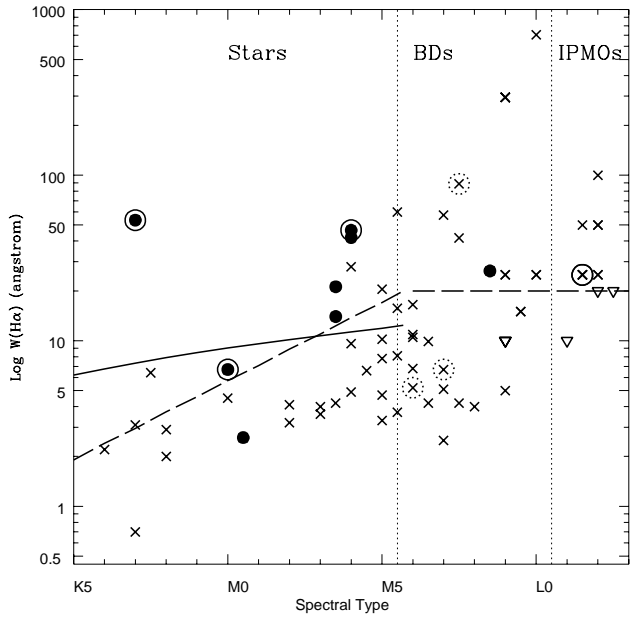}
 \caption{H$\alpha$ pseudo-equivalent width versus spectral type for
 $\sigma$\,Orionis low-mass members.  Crosses and open triangles
 correspond to data from
 this work, Barrado y Navascu\'es et al$.$ (2001),
 B\'ejar et al$.$ (1999) and Zapatero
 Osorio et al$.$ (2002a) --triangles denote the position of upper limits.
 Overlapping solid circles stand for objects with
 forbidden emission lines (Zapatero Osorio et al$.$ 2002a), whereas
 large open circle discriminate those cluster members with
 near-infrared excesses.  Broken open circles denote possible
 near-infrared excesses.
  The
 solid line separates CTT from WTT stars, whereas the long dashed line
 correspond to the upper envelope to atmospheric activity (see text).
 }
    \end{figure}
%______________________________________________________________      

%-----------------------------------------------------------
    \begin{figure}
    \centering
    \includegraphics[width=8.8cm]{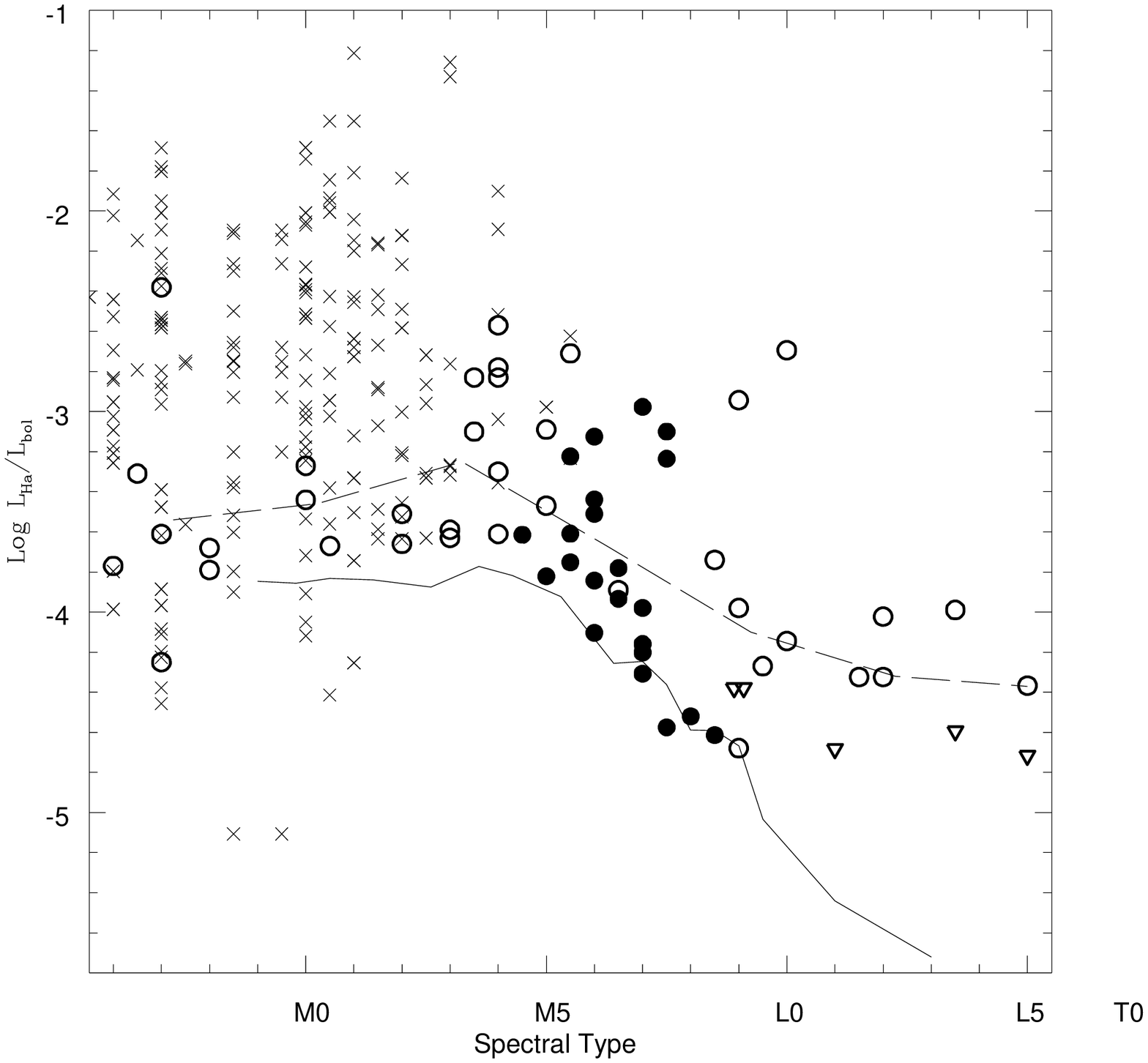}
 \caption{Ratio between the H$\alpha$ and bolometric luminosities
 against spectral type. Data for the Orion pre-main sequence
 population are shown with crosses, whereas circles and triangles
 (upper limits) represent $\sigma$\,Orionis members. The solid and
 long-dashed lines indicate the average locus of field stars 
(after Hawley, Gizis \& Reid 1996 and Gizis et al. 2000) and
 $\sigma$\,Orionis objects, respectively.  } 
     \end{figure}
%______________________________________________________________

\subsection{$K$-band infrared excesses}
So far, there is no agreed theoretical mechanism that can
satisfactorily explain the formation of brown dwarfs and
planetary-mass objects in isolation.
See Bodenheimer (1998), Pickett et al (2000),
Boss (2001), Reipurth (2002), Bate (2002),  
and references therein.
 The study of infrared excesses,
which could be ascribed to the presence of cool ``circumstellar''
disks, can shed new light on this topic. Figures 5a, 5b and 5c depict the
optical and 
near-infrared color-color diagram of $\sigma$\,Orionis very low-mass
stars and brown dwarfs. We note that planetary-mass cluster objects
are not included in this discussion because they lack $H$-band
photometry (they are too faint to be detected by 2MASS). 
Stars and brown dwarfs of $\sigma$ Orionis are displayed as circles
(solid circles from those showing forbidden emision lines, Zapatero 
Osorio et al. 2002a), whereas isolated planetary mass objects appear
as open triangles (in this last case, only in panel a). 
For few objects with large errors in the 2MASS photometry, 
we have displayed them using $IJK$ values from B\'ejar et al. (2001) and
$H$ from 2MASS. They are marked in the figures with the labels
``2M'' and ``B2001''.
As comparison, classical TTauri stars from Orion (Herbig \& Bell 1998) 
are displayed as crosses in panel a and b.
 The main sequence loci (Bessell \& Brett 1988;
Kirkpatrick et al$.$ 2000; Leggett et al. 2001) and the averaged 
T\,Tauri loci (this paper and Meyer et al. 1997)
are delineated by solid and dashed lines, respectively.
 Similar IR color-color diagrams have
been obtained by Tej et al$.$ (2002) and Oliveira et al$.$
(2002). These authors conclude that no significant near-infrared
excess is evident from the ($J-H$) versus ($H-K$) 
diagram, and that only
a fraction of about 6\%~of the $\sigma$\,Orionis members may be
affected by an excess in the $K$-band. From Figure~5, 
and taking into account the spectral types,
we infer that at
least four objects out of a total of 74 with $IJHK$ photometry,
 do show an excess at 2.2\,$\mu$m 
(namely  4771-41, r053840-0230, S\,Ori\,J054001.8 and  S\,Ori\,47).
Another three might have K excess too, but their
2MASS photometry has quite large uncertainties or their location in some
 of these diagrams is close to the main sequence loci
(i.e., S\,Ori\,J053912.8-022453, S\,Ori\,J053948.1-022914 and S\,Ori\,42). 
Thus,
based solely on $K$-band data and assuming that the flux excess is due
to the presence of disks surrounding the central object, we derive that
the disk frequency among the $\sigma$\,Orionis low-mass population is
in the range 5--9\%.
Note that one object, r053831-0235, has colors (very red) 
and a spectral type (M0)
which indicate that it is strongly reddened. Since it is a probable
 member of the cluster (it has lithium), 
this could be due to a small local cloud of dust around the star 
(in fact, it might have a  disk too). If reddened objects are included
in the computation, $\sigma$ Orionis cluster disk frecuency, based on optical 
and near-IR photometry, increases up to 12\%.

However, we remark that disk frequency as measured by $K$-band excess
can be underestimated. Jayawardhana et al$.$ (2002) obtained $L'$
photometry of six very low-mass cluster stars and brown dwarfs and
found one object, S\,Ori\,12, with significant $L'$ excess
(i.e., 16\%). S\,Ori\,12
has a mass estimated at the substellar limit. It does not display a
flux excess in the $K$-band data (as shown in Figure~5 as a large cross),
 which suggests that its surrounding disk is neither massive nor warm. It is
very likely that a higher fraction of low-mass stars and brown dwarfs
of the cluster possesses rather cool ``circumstellar'' material with
signatures that cannot be detected at near-infrared wavelengths.
Moreover, Fern\'andez \& Comer\'on (2001), by collecting 
optical and infrared
spectra and  photometry of LS-RCrA\,1, a star close to the 
substellar boundary, have shown that an object with 
very  strong emission lines  (permitted and forbidden alike)
does not have necessarily near infrared excesses at 2.2 microns.
On the other hand,  disks appear to be common in very young brown dwarfs
since many authors have reported evidences on their existence in
various star-forming regions (this paper, Muzerolle et al$.$ 2000;
Muench et al$.$ 2001; Natta \& Testi 2001; Mart\'\i n et al$.$ 2001;
Natta et al$.$ 2002; Jayawardhana et al$.$ 2002; Apai et al$.$ 2002;
Testi et al$.$ 2002). In addition, disk frequency among brown dwarfs
resembles that of very low-mass stars (Lada et al$.$ 2002). This may
indicate that brown dwarfs and low-mass stars share a common origin.
%Note that the unchanged shape of the mass
%function as one goes from the low-mass stellar regime to the
%substellar domain (e.g., Luhman et al$.$ 2000; Tej et al$.$ 2002;
%Jameson et al$.$ 2002; B\'ejar et al$.$ 2001).
 However, the precise
physical mechanism or mechanisms that give birth to free-floating
brown dwarfs and planetary-mass objects remain unknown, although it
seems that processes leading to the formation of disks around the
central nascent object are more likely.

%-----------------------------------------------------------
    \begin{figure*}
    \centering
    \includegraphics[width=7.2cm]{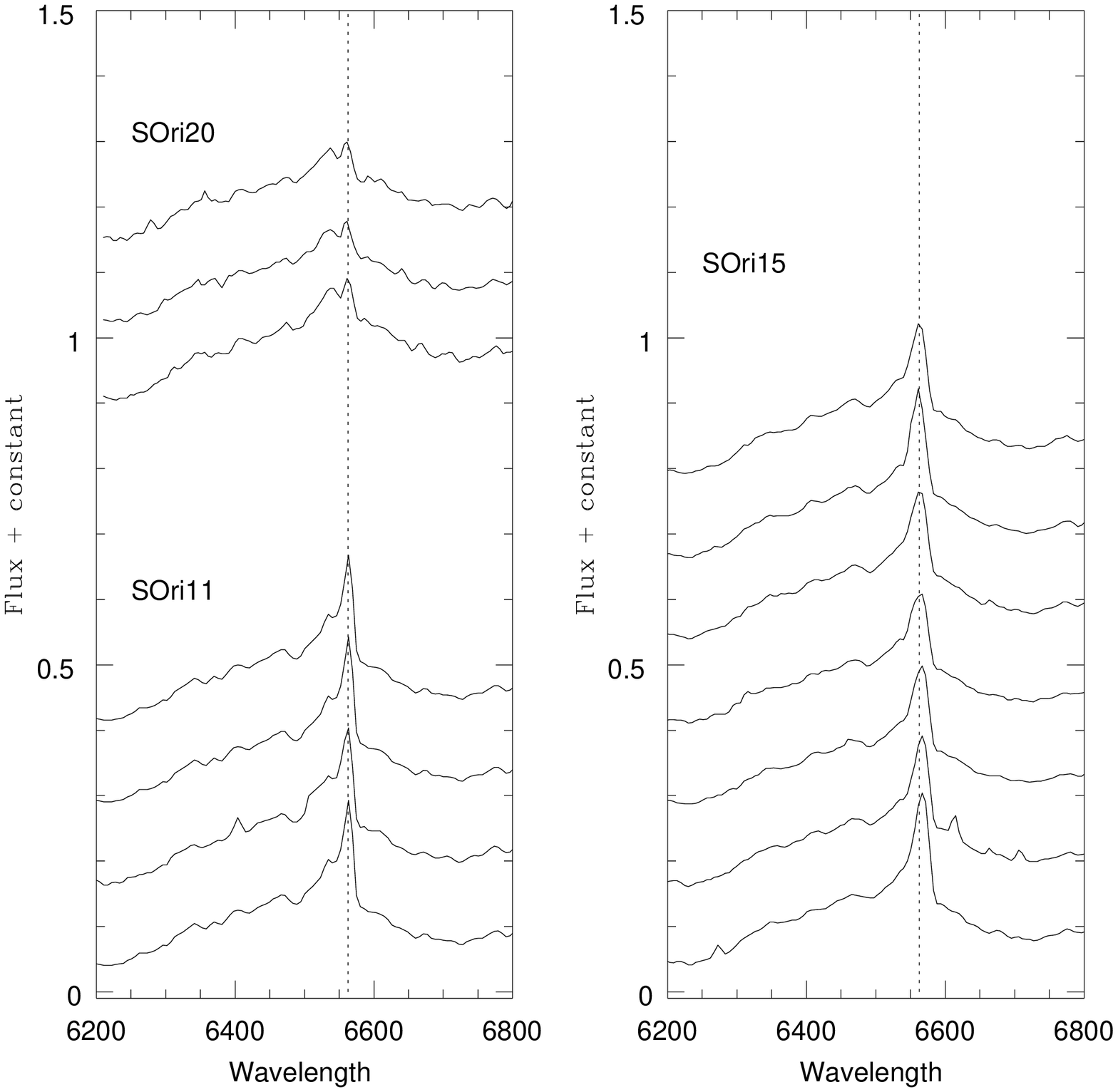}
    \includegraphics[width=7.2cm]{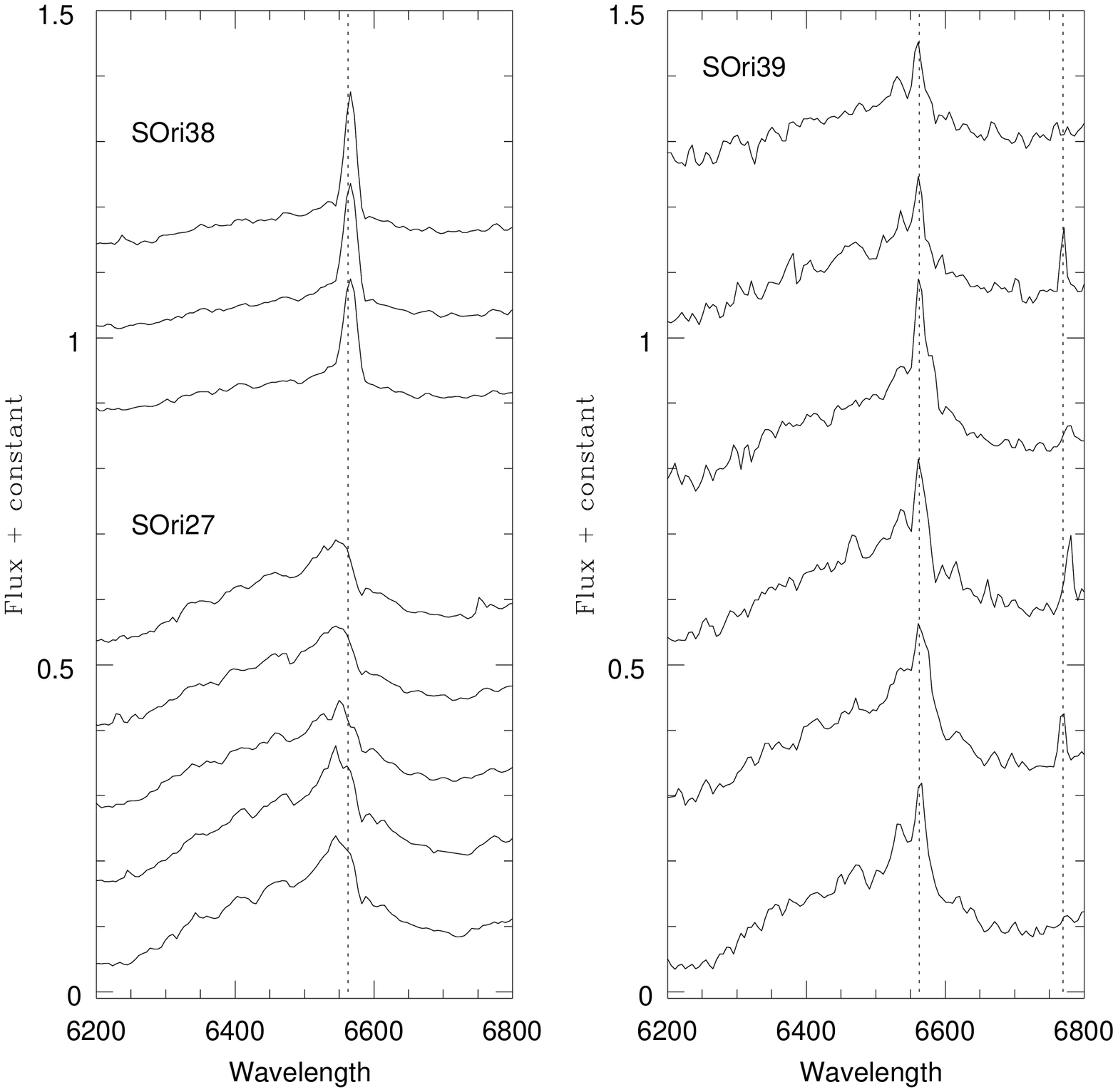}
    \includegraphics[width=7.2cm]{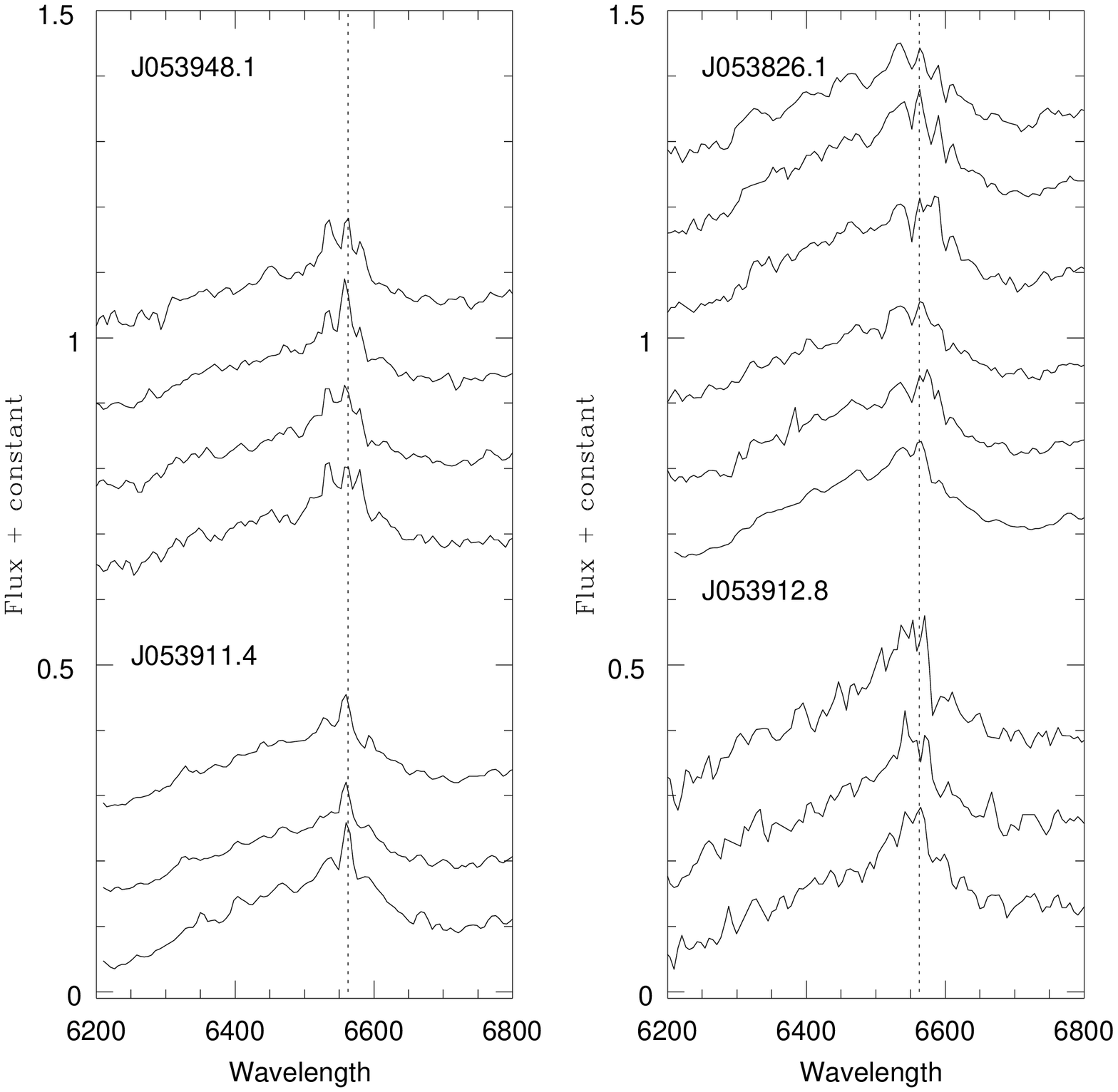}
 \caption{Individual VLT spectra showing in detail the area  around the H$\alpha$ line.  }
    \end{figure*}
%______________________________________________________________      

\subsection{H$\alpha$ 6563\AA{ } and its origin.}
%%%%%%%%%%%%%%%%%%%%%%%%%%%%%%%%%%%%%%%%%%%%%%%%%%

The equivalent width of the H$\alpha$ line at $\lambda$6563\AA{ } is
considered as an age indicator in M-dwarfs, and it is usually
associated to stellar activity. In general, the stronger the emission
line for a given spectral type, the younger the object. 
Intense and variable emission characterizes flare stars too 
(UV Cet type, see Gershberg et al. 1999).
Strong  H$\alpha$ emission also appears in episodes of accretion from
a close companion in interacting binaries or 
from a circumstellar disk, as happens in the TTauri stars and other
 young  stars.
In this case, the emission is produced as the accreted material
is channeled by magnetic field from the disrupted disk onto the
central object (Camenzind 1990), at nearly free-falling velocities.

We have identified  H$\alpha$ in emission  in all 
of the objects listed in Table~1,
 except in one case (S\,Ori\,J053909.9-022814) 
 and have  measured the pseudo-equivalent width (pW$_\lambda$),
since the spectral range around the H$\alpha$ contains a large 
number of intense molecular absorption lines.
This was carried out using the `splot' task and direct integration
within the IRAF environment.
 The results are
listed in column \#7 of Table~1. 
These values correspond to the average emission
when several spectra are available.
Individual measurements, exposure times and the modified Julian Date (MJD)
are listed in Table~3.
The spectral range around this feature is shown in
Figure 6, where we have included all the average spectra.
Note the very intense emission in the case of S\,Ori\,42, close to 
90 \AA, and  S\,Ori\,38, with 57 \AA, and S\,Ori\,25, with about 42 \AA.

Figure 7 illustrates  the behavior of the H$\alpha$ feature
--equivalent widths--  as a function
of the spectral  type. 
Solid circles represent the $\sigma$ Orionis 
candidate members analyzed in this paper, whereas
open circles correspond to members from our previous 
works in the cluster (B\'ejar et al. 1999;
Barrado y Navascu\'es et al. 2001;
Zapatero Osorio et al. 2002a).
Plus symbols and crosses represent the location
of pre-main sequence (classical TTauri stars, CTT) and 
weak-line TTauri stars
(WTT) stars from the Orion population, respectively
(Herbig \& Bell 1988 and Alcal\'a et al. 1996).
The  solid line
delimits  the areas corresponding to weak-line and classical TTauri stars.
For a comprenhensive discussion of how this criterium was defined,
see Barrado y Navascu\'es et al. (2003).
Finally, 
the dotted vertical segments separate the stellar, brown dwarf and 
planetary-mass domains.

CTT stars are characterized by their emission-line spectrum and 
non-photospheric continuum excesses, especially in the blue/UV and infrared.
Some forbidden emission lines are also present 
in some cases (see Appenzeller \& Mundt
1989 and Bertout 1989 for reviews). On the contrary, WTT stars
have spectra typical of main sequence stars, except
because the moderate H$\alpha$ emission and the 
strong LiI6708 \AA{ } doublet in absorption,
 clear indications of youth.
An easy, although not accurate, way of distinguishing 
whether a star belongs to one or the other category is the 
H$\alpha$ equivalent width: larger than 5--20 \AA{ } correspond to
CTT stars, and smaller to WTT. The actual criterion depends on different 
authors and the spectral types of the objects.
We have chosen a more elaborate criterion, 
based on data collected in several star forming regions
such as Orion, Taurus, Sco-Cen and Chamaleon
(see Barrado y Navascu\'es et al. 2003, which discusses
 an updated  version of the 
quantitative criteria defined in Mart\'{\i}n 1997).
From the physical point of view, the CTT star features are due 
to the presence of an accretion disk, which
induce, directly or indirectly, 
the formation of the intense H$\alpha$ emissions and other emission lines
(such as other lines from the Balmer series as well as  forbidden lines)
 and the  continuum excesses.   The H$\alpha$ emission is probably 
produced by the material which is being accreted,
whereas the hot continuum  comes from the areas where this material hits the 
central object. The IR excesses correspond to reprocessed
energy by the disk (see section 3.3).
In some cases there are hot disks which emit
large amount of energy at $\sim$2 $\mu$m
and a significant fraction of the total luminosity comes from the disk.
 Emission forbidden lines 
correspond to jets/outflows. Some blue-shifted absorption components
of permitted lines  are generated by strong winds coming from the inner
part of the disk, whereas red-shifted  absoptions are direct evidence
of infall. 
In the case of the WTT stars, 
the disk and all the phenomenology associated to
 it  have  disappeared, and the H$\alpha$ emission is interpreted as
enhanced solar-like chromospheric activity induced by rapid rotation.
This diagram seems to indicate that there are 
a  population of CTT stars in $\sigma$ Orionis.
In fact, the extrapolation of the WTT/CTT dividing line into 
the brown dwarf domain suggests that there are  
substellar counterparts to the CTT stars, or
 classical TTauri substellar  analogs (CTTSA).
Note that, due to its youth, all late spectral type members 
of the $\sigma$ Orionis cluster should have lithium.
Some measurements have been obtained by Zapatero Osorio et
al (2002a).

Figure 8 is similar to  Figure 7. In this case, 
we only plot $\sigma$ Orionis data (crosses and open triangles
 for upper limits).
Overlapping solid circles indicate those objects from
Zapatero Osorio et al. (2002a) which have forbidden emission lines,
whereas large open circles discriminate those members which have
IR excesses in the K band.
The solid line is the dividing line between WTT and CTT stars,
 whereas the long-dashed line represents the maximum H$\alpha$
equivalent width due to chromospheric activity (derived 
using data from young clusters such as IC2391, Alpha Per
and the Pleiades. See Barrado y Navascu\'es et al. 2003 for details).
This last curve indicates that most of  the $\sigma$ Orionis 
stellar and substellar population have pW$_\lambda$(H$\alpha$) in agreement
with chromospheric origin. However, some of them show a large
emission, which might have different source.
We emphasize that only the medium resolution spectra of
 low mass stars and BDs, discussed in  Zapatero Osorio et al. (2002a),  
are  good enough to detect the forbidden lines, and that 
these lines are not always present in CTT stars. 
The diagram shows a trend
between the detection of these lines, the presence of
IR excesses and strong H$\alpha$ emission measured
at this spectral resolution. These three phenomena indicate that,
at least, there are a handful of CTT stars and 
 substellar classical TTauri analogs in the
$\sigma$ Orionis cluster.
Moreover, as stated before, the VLT spectroscopic data have a low
 resolution which cannot allow the detection of emission
lines with few \AA, typical of forbidden lines 
which sometimes appear in the spectrum of CTT stars. 
However,  in the literature, the detection of lithium and the
H$\alpha$ emission have been extensively  used as criteria 
to catalog young  stars as either CTT or WTT stars.
Therefore, we can assume that some substellar objects
can be tentatively classified as 
classical TTauri analogs.

Table~4 summarized all the available information known so
far for low mass stars and BDs of the $\sigma$ Orionis cluster.
 Column \#3 indicates whether the object has IR excess
(based on optical and $JHK$ 2MASS data).
 Column \#4 states whether
the pW$_\lambda$(H$\alpha$) is above or below the dividing lines 
between CTT and WTT stars and the upper limit
of chromospheric activity (see Figure 8).
 Column \#5 contains information regarding the presence of forbidden
lines, whereas
 column \#6 lists the detection of LiI6708 \AA, when available.
 Finally, our classification as 
classical or weak-line TTauri stellar
(or substellar analog) is shown in 
 Column \#7.
This classification was carried out in 
a hierarchical order and attending to the following 
scheme:\\
i) Those members with forbidden lines have been catalogued as
probable CTT, and flagged in the table with ``CTT+''.\\
ii) Four objects present probable K excess  (``CTT?'' flag).\\
iii) Three BDs have possible IR excesses, 
although the 2MASS errors are large or their location is close to the
main sequence loci in some color-color diagram (``CTT??'' flag).\\
iv) Those objects with intense H$\alpha$ emission are
possible classical TTauri stars or substellar analogs
(``CTT-'' flag). In some cases, lithium has been detected. In others,
its presence is assumed due to the confirmed membership
and the age of  the cluster (1--8 Myr, Zapatero Osorio et al. 2002a).\\
v) Whenever lithium has been detected and the object
displays  a moderate  H$\alpha$ emission (below the 
criteria defined previously), we have catalogued it as weak-line TTauri
star or substellar analog (``WTT'' label).\\
vi) Finally, $\sigma$ Orionis members with moderate 
 H$\alpha$ emission and unknown lithium (due to the low
resolution spectroscopy) appear as possible weak-line TTauri
stars or analogs.
Note that the lack or an uncertain
 of detection of IR excess (or forbidden lines)
does not mean that they are not present, since additional
IR photometry in the 1-5 $\mu$m range and/or optical 
spectroscopy (better at higher resolution) with improved
signal--to--noise ration might provide positive detections
in some cases. An example is represented by
S\,Ori\,47. The 2MASS data indicate that
this BD close to the planetary domain and discussed deeply in 
Zapatero Osorio et al. (1999) and Barrado y Navascu\'es et al. (2001),  
might  have an infrared excess.
In this case, 2MASS errors are quite large.
Our UKIRT data corresponding to S\,Ori\,47, more accurate
than the 2MASS values, indicate that the disk is likely present.

Oliveira et al. (2002), by studying 
optical and infrared data of $\sigma$ Orionis candidates, were not able
to detect infrared excesses, which also characterized the 
presence of gas-dusty disks around stellar and substellar 
objects.   However, they do not provide a list with the results or
 the names of the objects, and we were not able to cross-correlate
their results with our BDs.
On the contrary,  Muench et al. (2001) 
have been able to detect some evidences for circumstellar disks
around BDs, down to 0.02 solar masses, 
in the Trapezium cluster ($\sim$1 Myr), 
by measuring  infrared excesses in about 65\% of the sample.
 Therefore, until further evidence is collected
(via higher signal-to-noise, higher resolution spectra), 
these data seem to indicate that  the H$\alpha$
weak emission comes from the photosphere of the objects, and could be 
analogous  to activity present in late spectral type stars
due to the chromosphere, except in a handful of the
$\sigma$ Orionis BDs, which show large 
equivalent widths, where the origin might be related to
accretion.

Figure 7 and Figure   8  clearly show that  the  H$\alpha$
emission tends to be  larger for cooler objects.
In the case of the PMS Orion population from the catalog by 
Herbig \& Bell (1988), there is a large increase in the strength
at about K0 spectral type. 
In the case of $\sigma$ Orionis cluster members,
the change in the behavior appears  at about M3.5 spectral type
(except in one case, 4771-41, a K7 star with large 
infrared excesses and characterized by forbidden lines,
 Zapatero Osorio et al. 2002a),
 when they are fully convective. 
Note, however, that part of this increase might be due to the
strengthening of TiO bands which deplete the continuum, and, therefore,
 enhance the H$\alpha$ equivalent width.
This  increase  might not be due to the flux drop in the continuum
in cooler objects.
In any case, a large range of values is present for a given 
spectral type in both groups.

We detect intense emission lines even 
for very low mass objects, such as in the case of S\,Ori\,55, which
has a mass $\sim$12 M$_{jupiter}$ and variable H$\alpha$ emission,
with pW$_\lambda$(H$\alpha$) between 185 and 410 \AA{} 
(Zapatero Osorio et al. 2002b),
and 5 \AA{ } (Barrado y Navascu\'es et al. 2001).
Another object with a mass close to 20 M$_{jupiter}$, S\,Ori\,71, has the
second largest emission  ever detected, as far as we know, in a star
or BD, with   pW$_\lambda$(H$\alpha$)=705 \AA{ } (Barrado y Navascu\'es et al.
2002a; Luhman et al. 2003).
 This spectral feature seems to be  asymmetric and very broad.
The source of these emissions is not clear.

One of the aims of the present paper
 is to try to establish the origin 
of the H$\alpha$ emission seen in most of the
 $\sigma$ Orionis low mass members (stars, BDs and IPMOs).
In this context, it is important to know how important is 
for the object the amount of energy release throughout the
line. We have computed the  ratio between the H$\alpha$ luminosity
(L$_{\rm H\alpha}$) to the bolometric luminosity (L$_{\rm bol}$).
The L$_{\rm H\alpha}$ and L$_{\rm bol}$ values were computed
following Herbst \& Miller (1989) and 
Hodgkin et al. (1995). See Zapatero Osorio et al. 
(2002a) for details. Figure 9  displays the ratio between 
L$_{\rm H\alpha}$ to L$_{\rm bol}$, 
versus the derived spectral type.
Members of the  $\sigma$ Orionis cluster are shown as open
and solid circles, as in Figure 3 and 4. 
The average locus for field stars and BDs is plotted as a solid line,
whereas the average ratio for $\sigma$ Orionis members is
displayed as a long-dashed line.
Clearly, most the of  $\sigma$ Orionis members have ratios higher
or much higher than field objects. In fact, most of the 
isolated planetary-mass objects and about half of the BDs
of  the $\sigma$ Orionis cluster have ratios about one order of magnitude
larger than their field objects counterparts. Note, however,
that field objects are much older and have larger masses.
For example, M5 and L0 from the field have masses of
 0.090 M$_\odot$ and 0.040  M$_\odot$  for an age of 100 Myr,
 0.110 M$_\odot$ and 0.075  M$_\odot$  for an age of 1 Gyr,
 whereas the same spectral types  would have
 0.070 M$_\odot$ and 0.014  M$_\odot$ for 3 Myr,
  the most likely age for  $\sigma$ Orionis cluster (Zapatero Osorio et al
2002a). In any case, some of the cluster members have ratios close to the
 saturation limit which appear in active late spectral type stars
(see, for instance, Stauffer et al. 1997).

%--------------------------------------------------<---------
    \begin{figure*}
    \centering
    \includegraphics[width=7.2cm]{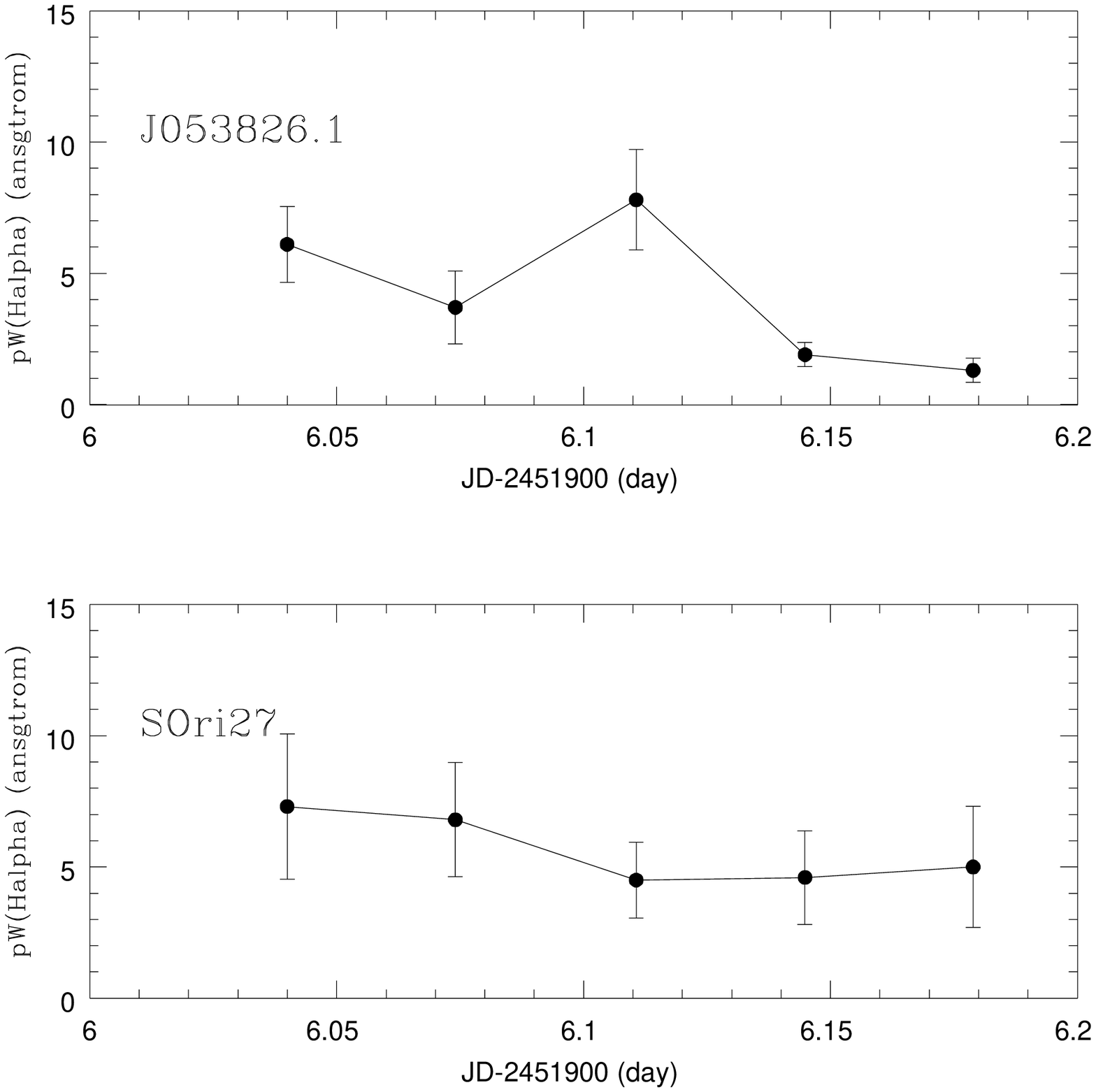}
    \includegraphics[width=7.2cm]{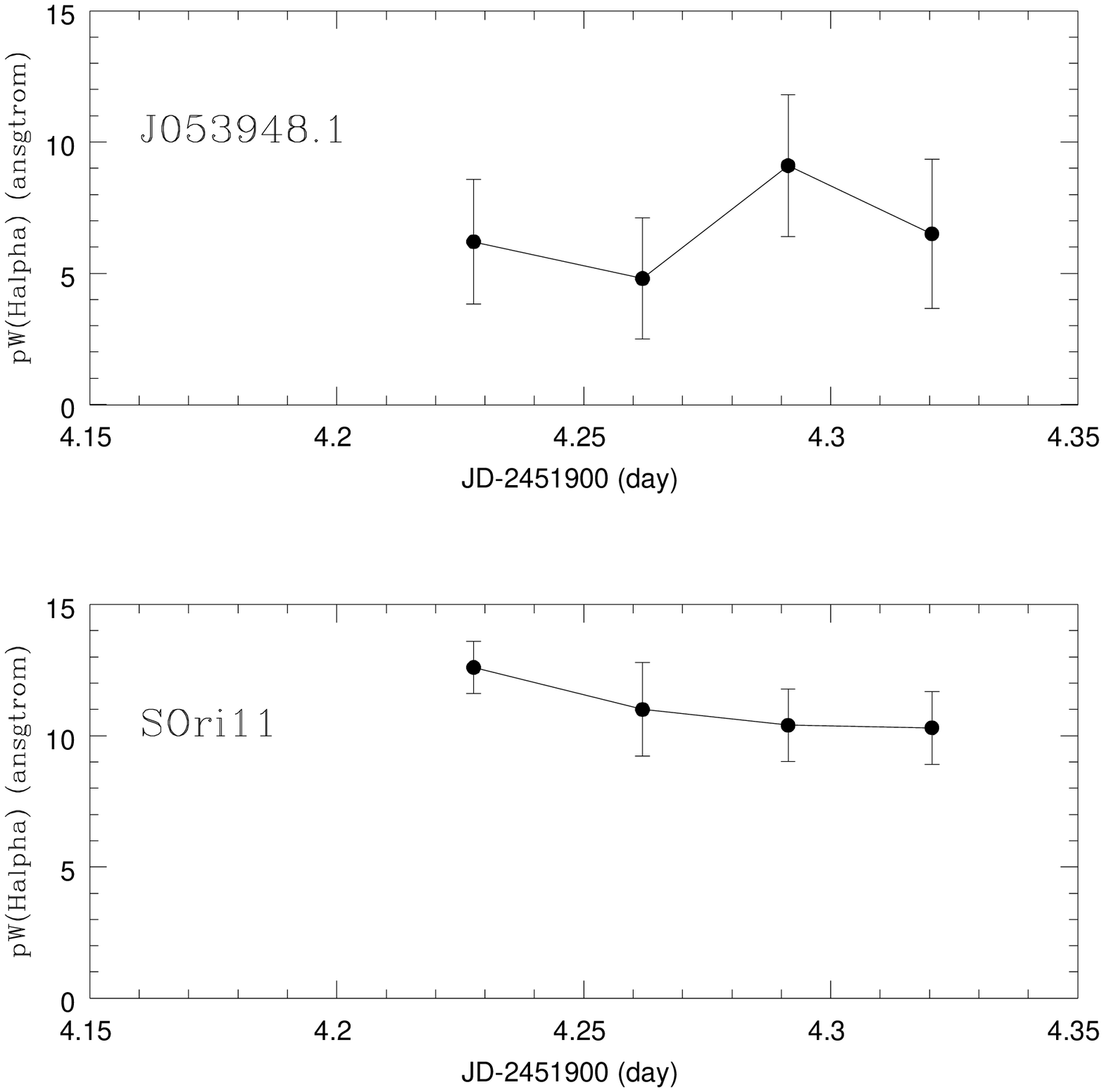}
    \includegraphics[width=7.2cm]{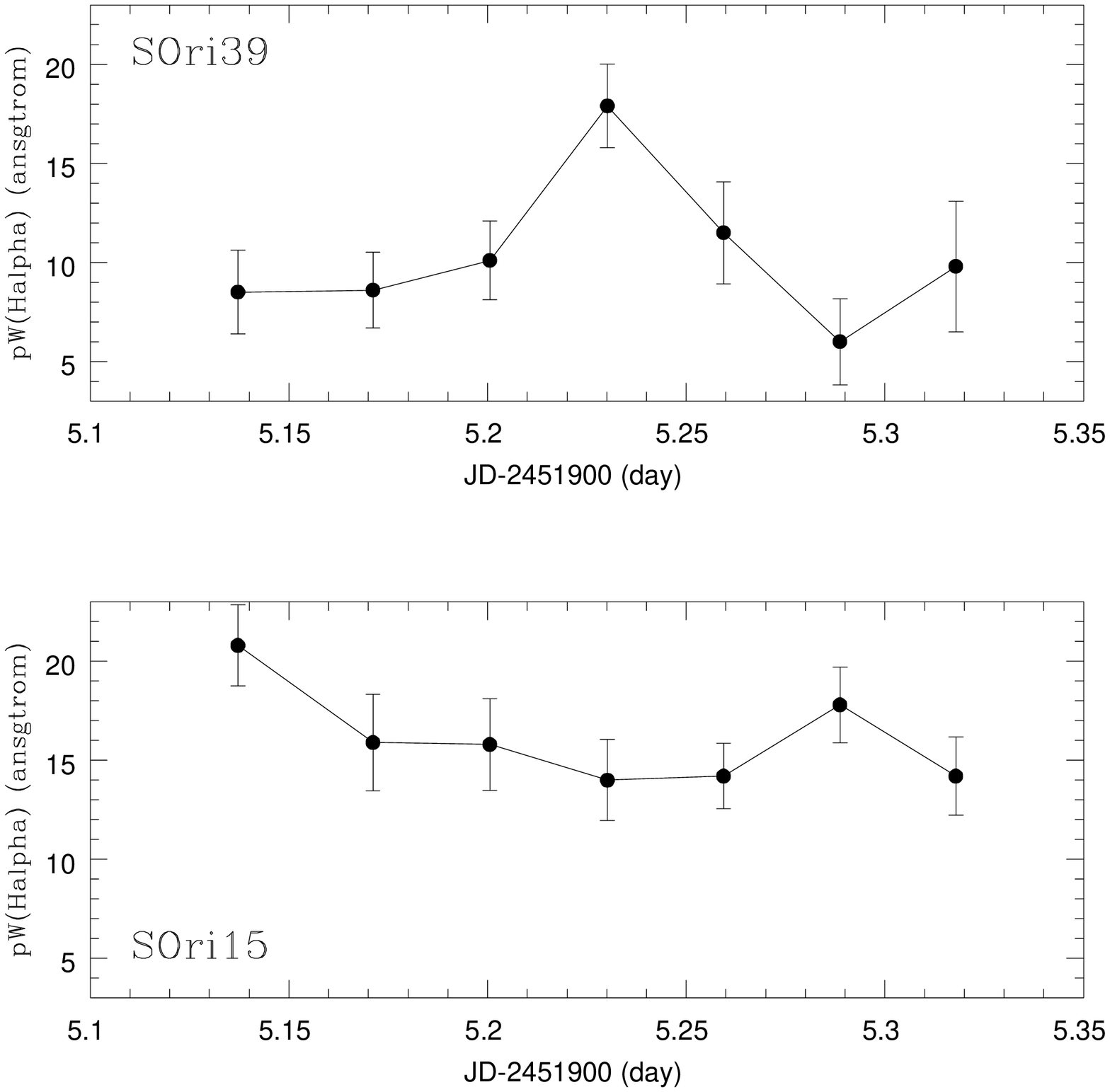}
 \caption{Pseudo H$\alpha$ equivalent with versus the
 Julian Date (minus 2451900) for several objects which have
 four or more independent spectra.}
          \label{}
    \end{figure*}
%______________________________________________________________      

\subsection{H$\alpha$ variability}

In addition to the large range of H$\alpha$ emission for a given 
spectral type, we have detected variability
(the most conspicuous cases being S\,Ori\,J053826.1-024041 and S\,Ori\,39).
This variability agrees with the WTT star classification for these
objects, since it means that its origin would be 
coronal/chromospheric and, therefore, intrinsically variable.
Note, however, that CTT stars have H$\alpha$ variability too, but in a 
larger time span or with sudden, un-modulated changes. 
Figure 10 shows the sequences of spectra around this feature
for several objects belonging to our  target list.
Although uncertainties are large for most cases, in several 
objects clear differences are present, as seen in the different  intensities.
If our interpretation of the   H$\alpha$  phenomenology is 
correct (see previous subsection), its variability might be due 
to the rotation of the objects, which is the ultimate origin of 
the chromospheric activity via the dynamo effect (Parker 1955). 
Bailer-Jones \& Mundt (2001), using photometric variability in the $I_c$
filter,  have measured some periodicities in several 
$\sigma$ Orionis members (S\,Ori\,31 and S\,Ori\,33) and failed to find them
in others (S\,Ori\,34, 44 and 46). This periods 
(7.5, 8.6 hours, respectively) might
correspond to the   rotation of the objects. Two of them are 
in  our survey. Unfortunately, they were only observed once.
In any case, it is likely that the   expected rotational
periods should be in the range of few hours, whereas
our exposure times were 40 minutes. Therefore, we should have covered 
a significant fraction of the phase, although
 the rotational periods are unknown.
None but one of the BDs with large H$\alpha$ emission line
(SOri\,38, tentatively classified as ``CTT-'' in Table 4)
 have more than one  spectrum and their variability was not investigated
for this reason.
Figure 11 shows the H$\alpha$ pseudo-equivalent widths versus
the Julian Date for all objects which have at least 4 
different measurements. 
Our time series cover about 3-4 hours,
half of the rotational periods measured in young BDs
belonging to the Pleiades and $\sigma$ Orionis clusters.
Some modulation might be present, specially in the case
of S\,Ori\,15 and S\,Ori\,39. A photometric campaign 
will unveil whether, indeed, they have strong 
photometric variability and whether their rotational periods
are in the range 2-4 hours, as the diagrams suggest.

\section{Conclusions}

We have collected low resolution spectroscopy of a sample
of low mass stars and brown dwarfs candidate members 
of the $\sigma$\,Orionis cluster and derived spectral types
by comparing with field objects. 
Infrared data from the 2MASS catalog was gathered for this
sample and other  $\sigma$\,Orionis candidate members.
Most of them 
are {\it bona fide} members of the cluster, based 
on the spectral type, the H$\alpha$ emission and the
 optical and infrared photometry.
The analysis of the infrared photometry indicates that
a small fraction (5--12\%) have K band excesses, a sign-post of 
dust disks. Moreover, by analyzing several properties 
(IR excess, intense  H$\alpha$ emission, 
detection of forbidden lines and lithium at 6708 \AA),
we have been able to classify a large number of the known
low mass candidate members of the cluster either as
classical or weak-line TTauri stars. Since 
few of them are, indeed, of substellar nature, 
a more proper name would be CTT or WTT substellar
analogs.
Finally, we have detected spectroscopic 
variability (in the H$\alpha$ equivalent width)
for some of them. The variability timespan is compatible 
with the expected rotational period of this type of objects.

\begin{acknowledgements}
We thank the  ESO staff at Paranal Observatory
and appreciate the excellent suggestions by
the referee, Xavier Delfosse.
 Partial  financial support was provided by the Spanish
This work has been partially financed by 
{\it ``Programa Ram\'on y Cajal''} 
and AYA2001-1124-CO2 programs.
EM acknowledges support from 
National Aeronautics and Space
Administration (NASA) grant NAG5-9992 and National Science Foundation (NSF)
grant AST-0205862. 
This research has made use of the NASA/IPAC Infrared Science
Archive, which is operated by the Jet Propulsion Laboratory,
California Institute of Technology, under contract with NASA.
\end{acknowledgements}

\end{document}

\baselineskip=16pt
\pagestyle{empty}
\small

\setcounter{table}{0}
\begin{table*}
\small
\caption[]{ Photometric and spectroscopic data for   $\sigma$ Orionis cluster
brown dwarf candidates.  }
\begin{tabular}{lll cc cc}
\hline
name        &Sp.Type& $I_c$   & $(R-I)_c$ & $(I_c-J)$ & alpha (J2000.0) delta&pW(H$\alpha$)  \\
            &    &            &           & &($^{h\,m\,s}$\hspace{5mm}$^\circ$ ' '')& (\AA)   \\
    (1)     &(2) &    (3)     &   (4)     &    (5)    &        (6)           &    (7)         \\
\hline                                                                           
SOri-J053909.9-022814    &M5&16.485 0.012& --    --  & 1.93 0.05 & 05 39 09.9 -02 28 14 &$<$3.3  --  \\ %so1-1-3199 c
SOri-J053911.4-023333    &M5&16.731 0.011& --    --  & 2.06 0.05 & 05 39 11.4 -02 33 33 &  4.7   1.0 \\ %so1-1-3113 d 
SOri-J053826.1-024041$^\ddagger$&M8&16.96 0.04& 2.04 0.06 & 2.02 0.07 & 05 38 26.1 -02 40 41 &  4.0   2.3 \\ % iac4-015   
SOri-J053829.0-024847    &M6&17.040 0.010& --    --  & 2.18 0.05 & 05 38 29.0 -02 48 47 &  --     -- \\ %so4-2-0731 e 
SOri-J053948.1-022914    &M7&18.921 0.009& 2.24 0.09 & 2.52 0.05 & 05 39 48.1 -02 29 14 &  6.7   1.6 \\ %iac5-041   b
SOri-J053912.8-022453    &M6&19.425 0.008& --    --  & 2.69 0.05 & 05 39 12.8 -02 24 53 &  5.2   2.3 \\ %so1-1-3055 f
SOri11        &M6      &16.424 0.008& 1.94 0.06 & 2.12 0.05 & 05 39 44.3 -02 33 01 & 10.5   1.5 \\ %   
SOri13        &M5.5    &16.410 0.018& 1.93 0.06 & 2.27 0.05 & 05 38 13.1 -02 24 10 &  8.1   2.0 \\ %  
SOri15        &M5.5    &16.789 0.014& 1.81 0.07 & 2.31 0.05 & 05 38 48.0 -02 28 54 & 15.7   2.3 \\ %  
SOri17$^1$    &M6      &16.945 0.009& 1.88 0.06 & 2.17 0.05 & 05 39 04.4 -02 38 35 & 10.9   2.7 \\ %  
SOri20$^+$    &M5.5    &17.321 0.009& 1.68 0.07 & 2.42 0.05 & 05 39 07.4 -02 29 08 &  3.7   1.0 \\ %  
SOri22        &M6      &17.109 0.008& 2.11 0.07 & 2.47 0.05 & 05 38 35.2 -02 25 24 &  6.8   3.4 \\ %  
SOri25$^2$    &M7.5    &17.163 0.008& 2.17 0.10 & 2.46 0.05 & 05 39 08.8 -02 39 58 & 41.8   8.5 \\ %  
SOri26$^*$    &M4.5    &17.264 0.008& 1.83 0.08 & 2.30 0.05 & 05 39 16.6 -02 38 27 &  6.6   3.5 \\ %  
SOri27$^3$    &M7      &17.08  0.05 & 2.13 0.07 & 2.22 0.05 & 05 38 17.3 -02 40 24 &  5.1   2.1 \\ %  
SOri29$^{+,4}$&M6.5    &17.230 0.008& 1.98 0.07 & 2.11 0.05 & 05 38 29.5 -02 25 17 &  4.2   1.5 \\ %  
SOri30        &M6      &17.438 0.008& 1.71 0.08 & 2.19 0.05 & 05 39 13.0 -02 37 51 & 16.5   5.5 \\ %  
SOri31$^a$    &M7      &17.429 0.008& 2.03 0.05 & 2.29 0.05 & 05 38 20.9 -02 46 13 &  2.5   0.9 \\ %  
SOri38        &M7      &17.640 0.008& 2.19 0.09 & 2.46 0.05 & 05 39 15.1 -02 21 52 & 57.4   7.9 \\ %  
SOri39$^5$    &M6.5    &17.922 0.008& 2.24 0.10 & 2.47 0.05 & 05 38 32.4 -02 29 58 &  9.9   3.7 \\ %  
SOri42        &M7.5    &19.01  0.09 & 2.47 0.41 & 2.33 0.20 & 05 39 23.3 -02 40 57 & 88.6  12.5 \\ %  
SOri44$^{b,6}$&M7      &19.39  0.06 & 2.31 0.15 & 2.12 0.06 & 05 38 07.0 -02 43 21 &  3.3   1.2 \\ %  
SOri45$^{7}$  &M8.5    &19.724 0.009& 2.75 0.017& 2.95 0.05 & 05 38 25.6 -02 48 36 & 26.4  15.4 \\ % $^c$ P(rot)=0.50$\pm$0.13 h  
SOri49$^*$    &M7.5    &20.83  0.07 &01.5   --  & 2.22 0.07 & 05 38 22.9 -02 37 55 &  4.2   3.7 \\ %  
\hline
\end{tabular}
$\,$\\
$^a$ P(rot)=7.5$\pm$0.6 hours, 
$^b$ Undetermined  P(rot) (Bailer-Jones et al. 2001).\\
Spectra types and previous pEW(H$\alpha$) from B\'ejar et al. (1999):
$^1$ M6 and 5.5 \AA, 
$^2$ M6.5 and 45.0 \AA, 
$^3$ M7 and 6.1 \AA,
$^4$ M6 and 28.0 \AA, 
$^5$ M6.5 and 5.1 \AA, 
$^6$ M6.5 and $<$5 \AA,
$^7$ M8.5 and 60.0 \AA\\
$^*$ Probable non-member based on the spectral type classification\\
$^+$ Probable non-member,  Kenyon et al. (2001). Based on a low S/N spectrum. \\ %--Cool Star Proceeding \\
$^\ddagger$ Photometric binary
\end{table*}

\setcounter{table}{1}
\begin{table*}
\tiny
\caption[]{Infrared  photometric  data, from 2MASS catalog,
 for all candidate members of  $\sigma$ Orionis cluster.
}
\begin{tabular}{l l c ccc cc l}
\hline
 Name               &MK  &  Ic   &J\,\,$\sigma$(J)& H\,\, $\sigma$(H)& Ks\,\, $\sigma$(K)& radius  &  RA (2MASS)  DEC     & Source \\
                    &    & (mag) &    (mag)       & (mag)            &     (mag)         & (arcsec)&($^{h\,m\,s}$\hspace{5mm}$^\circ$ ' '') & \\
\hline                                    
4771-1038           &K8  & 12.78 &  11.655 0.028 &  10.972 0.028 &  10.750 0.027 & 2.037 & 05 39 11.63  -02 36 02.8  & Wo1996,ZO2002  \\  % 
4771-1051           &K7.5& 12.33 &  -- \,\,\, -- &  -- \,\,\, -- &  -- \,\,\, -- & --    & -- \,\,\, \,\,\, --       & Wo1996,ZO2002  \\  % 
4771-1075           &K7  & 12.66 &  11.566 0.028 &  10.877 0.027 &  10.678 0.027 & 1.728 & 05 39 05.41  -02 32 30.2  & Wo1996,ZO2002  \\  % 
4771-1097           &K6  & 12.43 &  -- \,\,\, -- &  -- \,\,\, -- &  -- \,\,\, -- & --    & -- \,\,\, \,\,\, --       & Wo1996,ZO2002  \\  % 
4771-41             &K7  & 12.82 &  11.952 0.029 &  10.783 0.028 &   9.925 0.029 & 2.295 & 05 38 27.25  -02 45 09.7  & Wo1996,ZO2002  \\  % 
4771-899            &K7  & 12.08 &  10.168 0.027 &   9.478 0.029 &   9.194 0.027 & 2.055 & 05 38 48.04  -02 27 14.1  & Wo1996,ZO2002  \\  % 
r053820-0234        &M4  & 14.58 &  -- \,\,\, -- &  -- \,\,\, -- &  -- \,\,\, -- & --    & -- \,\,\, \,\,\, --       & Wo1996,ZO2002  \\  % 
r053831-0235        &M0  & 13.52 &  11.520 0.029 &  10.708 0.027 &  10.352 0.028 & 2.723 & 05 38 31.58  -02 35 14.8  & Wo1996,ZO2002  \\  % 
r053833-0236        &M3.5& 13.71 &  -- \,\,\, -- &  -- \,\,\, -- &  -- \,\,\, -- & --    & -- \,\,\, \,\,\, --       & Wo1996,ZO2002  \\  % 
r053838-0236        &K8  & 12.37 &  11.186 0.028 &  10.482 0.027 &  10.313 0.028 & 3.450 & 05 38 38.23  -02 36 38.4  & Wo1996,ZO2002  \\  % 
r053840-0230        &M0  & 12.80 &  11.540 0.028 &  10.781 0.028 &  10.414 0.028 & 1.267 & 05 38 40.28  -02 30 18.5  & Wo1996,ZO2002  \\  % 
r053849-0238        &M0.5& 12.88 &  11.407 0.027 &  10.683 0.028 &  10.504 0.029 & 2.899 & 05 38 49.18  -02 38 22.2  & Wo1996,ZO2002  \\  % 
r053907-0228        &M3  & 14.33 &  12.893 0.029 &  12.160 0.027 &  11.974 0.031 & 0.273 & 05 39 07.60  -02 28 23.3  & Wo1996,ZO2002  \\  % 
r053923-0233        &M2  & 14.16 &  12.835 0.029 &  12.120 0.028 &  11.863 0.031 & 2.578 & 05 39 22.87  -02 33 33.0  & Wo1996,ZO2002  \\  % 
SOriJ053715.1-024202&M4  & 15.07 &  13.675 0.033 &  13.021 0.040 &  12.688 0.040 & 0.925 & 05 37 15.16  -02 42 01.6  & Be2001,ZO2002  \\  % 
SOriJ053746.6-024328&    & 16.51 &  -- \,\,\, -- &  -- \,\,\, -- &  -- \,\,\, -- & --    & -- \,\,\, \,\,\, --       & Be2001         \\  %  
SOriJ053820.1-023802&M4  & 14.41 &  12.596 0.029 &  11.870 0.028 &  11.632 0.031 & 1.749 & 05 38 20.21  -02 38 01.7  & Be2001,ZO2002,$\dagger$  \\  % 
SOriJ053821.3-023336&    & 17.70 &  15.344 0.044 &  14.796 0.050 &  14.505 0.089 & 1.260 & 05 38 21.38  -02 33 36.3  & Be2001         \\  %
SOriJ053826.1-024041&M8  & 16.95 &  14.888 0.038 &  14.275 0.038 &  13.890 0.058 & 1.307 & 05 38 26.23  -02 40 41.4  & Be2001         \\  %soIAC4-015       
SOriJ053827.4-023504&M3.5& 14.50 &  12.815 0.030 &  12.126 0.027 &  11.835 0.030 & 1.567 & 05 38 27.50  -02 35 04.2  & Be2001,ZO2002  \\  % 
SOriJ053829.0-024847&M6  & 17.04 &  14.812 0.038 &  14.280 0.039 &  13.866 0.055 & 0.798 & 05 38 28.97  -02 48 47.3  & Be2001         \\  %so4-2-0731       
SOriJ053847.5-022711&M5  & 14.46 &  12.150 0.027 &  11.509 0.027 &  11.298 0.029 & 1.142 & 05 38 47.55  -02 27 11.9  & Be2001,ZO2002  \\  % 
SOriJ053849.2-022358&    & 16.81 &  14.375 0.032 &  13.695 0.033 &  13.214 0.038 & 1.419 & 05 38 49.29  -02 23 57.5  & Be2001         \\  %
SOriJ053909.9-022814&M5  & 16.49 &  14.620 0.036 &  14.015 0.035 &  13.802 0.050 & 3.064 & 05 39 10.02  -02 28 11.5  & Be2001         \\  %
SOriJ053911.4-023333&M5  & 16.73 &  14.481 0.033 &  13.944 0.035 &  13.574 0.044 & 0.337 & 05 39 11.40  -02 33 32.7  & Be2001         \\  %
SOriJ053912.8-022453&M6  & 19.42 &  16.848 0.133 &  15.728 0.106 &  15.202 0.163 & 1.347 & 05 39 12.90  -02 24 53.6  & Be2001         \\  %so1-1-3055       
SOriJ053914.5-022834&M3.5& 14.75 &  13.357 0.030 &  12.657 0.029 &  12.343 0.032 & 0.864 & 05 39 14.47  -02 28 33.2  & Be2001,ZO2002  \\  % 
SOriJ053920.5-022737&M2  & 13.51 &  12.157 0.028 &  11.433 0.028 &  11.174 0.029 & 0.861 & 05 39 20.45  -02 27 36.7  & Be2001,ZO2002  \\  % 
SOriJ053926.8-022614&    & 18.66 &  16.257 0.088 &  15.585 0.089 &  15.127 0.181 & 0.237 & 05 39 26.79  -02 26 13.9  & Be2001         \\  %
SOriJ053936.4-023626&    & 18.46 &  -- \,\,\, -- &  -- \,\,\, -- &  -- \,\,\, -- & --    & -- \,\,\, \,\,\, --       & Be2001         \\  %
SOriJ053946.5-022423&    & 20.14 &  17.045 0.170 &  16.152 0.136 &  15.398 0.000 & 0.745 & 05 39 46.45  -02 24 23.1  & Be2001         \\  %
SOriJ053948.1-022914&M7  & 18.76 &  16.445 0.094 &  15.599 0.088 &  15.165 0.147 & 1.806 & 05 39 48.26  -02 29 14.5  & Be2001         \\  %soIAC5-041       
SOriJ053949.3-022346&M4  & 15.14 &  13.437 0.029 &  12.753 0.029 &  12.445 0.035 & 2.135 & 05 39 49.44  -02 23 45.9  & Be2001,ZO2002  \\  % 
SOriJ053951.6-022248&M5.5& 14.59 &  12.617 0.028 &  12.001 0.028 &  11.674 0.030 & 2.054 & 05 39 51.73  -02 22 47.2  & Be2001,ZO2002  \\  % 
SOriJ053958.1-022619&M3  & 14.19 &  12.788 0.028 &  12.061 0.027 &  11.847 0.030 & 2.399 & 05 39 58.26  -02 26 18.8  & Be2001,ZO2002  \\  % 
SOriJ054001.8-022133&M4  & 14.32 &  12.374 0.028 &  11.588 0.027 &  11.264 0.028 & 2.471 & 05 40 01.96  -02 21 32.5  & Be2001,ZO2002  \\  % 
SOriJ054005.1-023052&M5  & 15.90 &  13.974 0.029 &  13.382 0.031 &  13.075 0.034 & 2.408 & 05 40 05.26  -02 30 52.2  & Be2001,ZO2002  \\  % 
SOri01              &    & 15.08 &  13.628 0.030 &  12.986 0.029 &  12.653 0.033 & 2.099 & 05 39 11.84  -02 27 40.8  & Be1999         \\  %
SOri02              &    & 15.12 &  13.521 0.030 &  12.852 0.029 &  12.579 0.033 & 2.086 & 05 39 26.34  -02 28 37.7  & Be1999         \\  %
SOri03              &    & 15.16 &  13.290 0.028 &  12.758 0.029 &  12.448 0.034 & 3.082 & 05 39 20.97  -02 30 33.3  & Be1999         \\  %
SOri04              &    & 15.23 &  13.466 0.029 &  12.910 0.029 &  12.556 0.033 & 2.504 & 05 39 39.32  -02 32 25.3  & Be1999         \\  %
SOri05              &    & 15.40 &  13.624 0.030 &  13.050 0.029 &  12.765 0.032 & 2.103 & 05 39 20.24  -02 38 25.8  & Be1999         \\  %
SOri06              &    & 15.53 &  13.474 0.029 &  12.846 0.028 &  12.590 0.031 & 2.503 & 05 38 47.66  -02 30 37.3  & Be1999         \\  %
SOri07              &    & 15.63 &  13.831 0.030 &  13.260 0.031 &  12.923 0.033 & 2.531 & 05 39 08.22  -02 32 28.3  & Be1999         \\  %
SOri08              &    & 15.74 &  14.154 0.031 &  13.520 0.032 &  13.276 0.039 & 4.432 & 05 39 08.09  -02 28 44.7  & Be1999         \\  %
SOri09              &    & 15.81 &  13.629 0.030 &  13.048 0.029 &  12.746 0.032 & 2.642 & 05 38 17.18  -02 22 25.7  & Be1999         \\  %
SOri10              &    & 16.08 &  14.201 0.033 &  13.549 0.031 &  13.174 0.037 & 2.392 & 05 39 44.50  -02 24 43.2  & Be1999         \\  %
SOri11              &M6  & 16.15 &  14.303 0.031 &  13.720 0.033 &  13.379 0.041 & 2.958 & 05 39 44.33  -02 33 02.8  & Be1999         \\  %
SOri12              &M6  & 16.28 &  14.200 0.033 &  13.636 0.032 &  13.279 0.040 & 0.959 & 05 37 57.45  -02 38 44.5  & Be1999         \\  %
SOri13              &M5.5& 16.35 &  14.231 0.033 &  13.574 0.032 &  13.262 0.037 & 3.001 & 05 38 13.21  -02 24 07.5  & Be1999         \\  %
SOri14              &    & 16.35 &  14.393 0.033 &  13.820 0.034 &  13.441 0.039 & 1.559 & 05 39 37.60  -02 44 30.5  & Be1999         \\  %
SOri15              &M5.5& 16.55 &  14.501 0.034 &  13.845 0.032 &  13.443 0.041 & 1.564 & 05 38 48.10  -02 28 53.6  & Be1999         \\  %
SOri16              &    & 16.60 &  14.670 0.035 &  14.057 0.035 &  13.652 0.044 & 1.684 & 05 39 15.11  -02 40 47.5  & Be1999         \\  %
SOri17              &M6  & 16.80 &  14.766 0.036 &  14.188 0.039 &  13.792 0.047 & 1.485 & 05 39 04.50  -02 38 35.3  & Be1999         \\  %
SOri18              &    & 16.81 &  14.651 0.038 &  14.057 0.039 &  13.821 0.059 & 1.196 & 05 38 25.68  -02 31 21.7  & Be1999         \\  %
SOri19              &    & 16.81 &  14.780 0.048 &  14.160 0.055 &  13.900 0.066 & 3.055 & 05 37 21.06  -02 25 40.1  & Be1999         \\  %
SOri20              &M5.5& 16.88 &  14.949 0.038 &  14.349 0.040 &  13.920 0.055 & 3.792 & 05 39 07.59  -02 29 05.5  & Be1999         \\  %
SOri21              &    & 16.90 &  14.787 0.036 &  14.197 0.040 &  13.837 0.054 & 1.991 & 05 39 34.33  -02 38 46.9  & Be1999         \\  %
SOri22              &M6  & 16.98 &  -- \,\,\, -- &  -- \,\,\, -- &  -- \,\,\, -- & --    & -- \,\,\, \,\,\, --       & Be1999         \\  %
SOri23              &    & 17.00 &  14.928 0.039 &  14.385 0.041 &  13.931 0.060 & 3.032 & 05 37 51.11  -02 26 07.4  & Be1999         \\  %
SOri24              &    & 17.00 &  14.993 0.039 &  14.412 0.046 &  14.122 0.064 & 2.189 & 05 37 55.74  -02 24 33.6  & Be1999         \\  %
SOri25              &M7.5& 17.04 &  14.665 0.035 &  14.151 0.035 &  13.756 0.051 & 2.362 & 05 39 08.96  -02 39 57.9  & Be1999         \\  %
SOri26              &M4.5& 17.05 &  -- \,\,\, -- &  -- \,\,\, -- &  -- \,\,\, -- & --    & -- \,\,\, \,\,\, --       & Be1999         \\  %
SOri27              &M7  & 17.07 &  14.845 0.037 &  14.329 0.038 &  14.097 0.063 & 1.801 & 05 38 17.42  -02 40 24.3  & Be1999,ZO2002  \\  %
SOri28              &    & 17.11 &  15.332 0.046 &  14.787 0.046 &  14.343 0.071 & 1.406 & 05 39 23.19  -02 46 55.9  & Be1999         \\  %
SOri29              &M6.5& 17.18 &  14.829 0.037 &  14.297 0.040 &  13.955 0.061 & 3.362 & 05 38 29.62  -02 25 14.1  & Be1999         \\  %
SOri30              &M6  & 17.25 &  15.270 0.042 &  14.774 0.050 &  14.288 0.073 & 1.324 & 05 39 13.09  -02 37 50.8  & Be1999         \\  %
SOri31              &M7  & 17.29 &  15.181 0.041 &  14.562 0.045 &  14.203 0.075 & 1.531 & 05 38 20.88  -02 46 13.3  & Be1999         \\  %
SOri32              &    & 17.36 &  15.346 0.050 &  14.785 0.051 &  14.370 0.077 & 1.553 & 05 39 43.59  -02 47 31.9  & Be1999         \\  %
SOri33              &    & 17.38 &  15.169 0.055 &  14.529 0.063 &  14.088 0.069 & 3.520 & 05 36 58.07  -02 35 19.5  & Be1999         \\  %
SOri34              &    & 17.46 &  15.516 0.066 &  15.082 0.098 &  14.622 0.119 & 2.422 & 05 37 07.22  -02 32 44.3  & Be1999         \\  %
SOri35              &    & 17.50 &  15.205 0.042 &  14.603 0.046 &  14.163 0.062 & 3.053 & 05 37 55.60  -02 33 05.3  & Be1999         \\  %
SOri36              &    & 17.67 &  15.469 0.048 &  14.844 0.052 &  14.550 0.076 & 0.865 & 05 39 26.86  -02 36 56.2  & Be1999         \\  %
SOri37              &    & 17.70 &  15.575 0.075 &  14.872 0.080 &  14.434 0.100 & 2.594 & 05 37 07.61  -02 27 15.0  & Be1999         \\  %
SOri38              &M7  & 17.70 &  15.598 0.051 &  14.938 0.057 &  14.573 0.090 & 2.880 & 05 39 15.26  -02 21 50.5  & Be1999         \\  %
SOri39              &M6.5& 17.82 &  15.465 0.047 &  14.854 0.051 &  14.430 0.082 & 1.033 & 05 38 32.45  -02 29 57.2  & Be1999         \\  %
SOri40              &M7  & 17.93 &  15.485 0.051 &  14.943 0.056 &  14.585 0.096 & 1.332 & 05 37 36.49  -02 41 56.7  & Be1999         \\  %
SOri41              &    & 18.44 &  16.728 0.120 &  15.737 0.096 &  15.717 0.229 & 3.212 & 05 39 38.50  -02 31 13.1  & Be1999         \\  %
SOri42              &M7.5& 19.01 &  16.724 0.130 &  15.897 0.115 &  15.589 0.209 & 1.755 & 05 39 23.41  -02 40 57.6  & Be1999         \\  %
SOri43              &    & 19.05 &  16.862 0.144 &  16.080 0.146 &  15.574 0.000 & 3.684 & 05 38 13.96  -02 35 01.2  & Be1999         \\  %
SOri44              &M77 & 19.39 &  -- \,\,\, -- &  -- \,\,\, -- &  -- \,\,\, -- & --    & -- \,\,\, \,\,\, --       & Be1999         \\  %
SOri45              &M8.5& 19.59 &  16.679 0.121 &  16.031 0.141 &  15.660 0.311 & 1.538 & 05 38 25.58  -02 48 37.0  & Be1999,ZO2002  \\  % 
SOri46              &    & 19.82 &  -- \,\,\, -- &  -- \,\,\, -- &  -- \,\,\, -- & --    & -- \,\,\, \,\,\, --       & Be1999         \\  %
SOri71              &L0  & 20.02 &  -- \,\,\, -- &  -- \,\,\, -- &  -- \,\,\, -- & --    & -- \,\,\, \,\,\, --       & --             \\  %sok05-0586       
SOri47(2MASS)       &L1.5& 20.53 &  17.519 0.251 &  16.250 0.149 &  15.807 0.283 & 2.086 & 05 38 14.63  -02 40 15.4  & ZO1999        \\  %
SOri47(UKIRT)       &    &  ''   &  17.305 0.027 &  16.458 0.032 &  15.806 0.028 &  ''   &           ''              & ZO1999        \\  %
SOri48              &    & 20.70 &  -- \,\,\, -- &  -- \,\,\, -- &  -- \,\,\, -- & --    & -- \,\,\, \,\,\, --       & Be2001        \\  %
SOri49              &M7.5& 20.83 &  -- \,\,\, -- &  -- \,\,\, -- &  -- \,\,\, -- & --    & -- \,\,\, \,\,\, --       & Be2001        \\  %
\hline
\end{tabular}
$\,$ \\
Wo1996 = Wolk (1996), Be1999 = B\'ejar et al. (1999), Be2001 = B\'ejar et al. (2001), 
ZO1999 = Zapatero Osorio et al. (1999), ZO2000 = Zapatero Osorio et al. (2000),
ZO2002 = Zapatero Osorio et al. (2002),
Ba2000 = Barrado y Navascu\'es et al. (2001).\\
$\dagger$ Same object as r053820-0237.\\
\end{table*}

\setcounter{table}{2}
\begin{table*}
\small
\caption[]{H$\alpha$ equivalent widths and observing dates for the  $\sigma$ Orionis cluster
brown dwarf candidates.  }
\begin{tabular}{llrl}
\hline
name    &     MJD-51000      & pW(H$\alpha$)& Exp.Time \\   %    
        &    (days)          &  (\AA)       & (seconds)\\   %
    (1) &      (2)           &  (3)         &    (4)                      \\   %
\hline      
% SOri16 &  904.05650249        &   --    --  & 2400  \\   %SOri50 VLT2000.00.N2.fits 
                                                                                 
 SOri11 &  904.22764064        &   12.6  1.5 & 2400  \\   %SOri11 VLT2000.04.N2.fits        
   ''   &  904.26190299        &   11.0  2.7 & 2400  \\   %SOri11 VLT2000.05.N2.fits        
   ''   &  904.29136284        &   10.4  2.1 & 2400  \\   %SOri11 VLT2000.06.N2.fits        
   ''   &  904.32053636        &   10.3  2.1 & 2400  \\   %SOri11 VLT2000.07.N2.fits        
                                                                                 
 SOri13 &  903.32829829        &    8.1  2.0 & 2580  \\   %SOri13 VLT2000.07.N1.fits        
                                                                                 
 SOri15 &  905.13713833        &   20.8  3.1 & 2400  \\   %SOri15 VLT2000.02.N3.fits        
   ''   &  905.17120378        &   15.9  3.7 & 2400  \\   %SOri15 VLT2000.03.N3.fits        
   ''   &  905.20057526        &   15.8  3.5 & 2400  \\   %SOri15 VLT2000.04.N3.fits        
   ''   &  905.23010044        &   14.0  3.1 & 2400  \\   %SOri15 VLT2000.05.N3.fits        
   ''   &  905.25940527        &   14.2  2.5 & 2400  \\   %SOri15 VLT2000.06.N3.fits        
   ''   &  905.28869611        &   17.8  2.9 & 2400  \\   %SOri15 VLT2000.07.N3.fits        
   ''   &  905.31784490        &   14.2  3.0 & 2400  \\   %SOri15 VLT2000.08.N3.fits        
                                                                                 
 SOri17 &  904.05650249        &   10.9  2.7 & 2400  \\   %SOri50 VLT2000.00.N2.fits 
                                                                                 
 SOri20 &  903.10135488        &    3.2  0.9 & 2400  \\   %SOri20 VLT2000.02.N1.fits        
   ''   &  903.14482793        &    2.9  0.8 & 2400  \\   %SOri20 VLT2000.03.N1.fits        
   ''   &  903.17441719        &    3.3  1.2 & 2400  \\   %SOri20 VLT2000.04.N1.fits        
                                                                                 
 SOri22 &  905.06892540        &    6.8  3.4 & 4800  \\   %SOri22 VLT2000.00.N3.fits        
                                                                                 
 SOri25 &  904.05650249        &   41.8  8.5 & 2400  \\   %SOri50 VLT2000.00.N2.fits 
                                                                                 
 SOri26 &  904.05650249        &    6.6  3.5 & 2400  \\   %SOri50 VLT2000.00.N2.fits 
                                                                                 
 SOri27 &  903.05138618        &    5.5  3.1 & 2400  \\   %SOri47 VLT2000.00.N1.fits 
   ''   &  906.03996572        &    7.3  4.2 & 2400  \\   %SOri27 VLT2000.00.N4.fits        
   ''   &  906.07400201        &    6.8  3.3 & 2400  \\   %SOri27 VLT2000.01.N4.fits        
   ''   &  906.11072522        &    4.5  2.2 & 2400  \\   %SOri27 VLT2000.02.N4.fits        
   ''   &  906.14487566        &    4.6  2.7 & 2400  \\   %SOri27 VLT2000.03.N4.fits        
   ''   &  906.17892987        &    5.0  3.5 & 2400  \\   %SOri27 VLT2000.04.N4.fits        
                                                                                 
 SOri29 &  905.06892540        &    4.2  1.5 & 4800  \\   %SOri29 VLT2000.00.N3.fits        
                                                                                 
 SOri30 &  904.05650249        &   16.5  5.5 & 2400  \\   %SOri50 VLT2000.00.N2.fits 
                                                                                 
 SOri31 &  903.23775122        &    2.5  0.9 & 4800  \\   %SOri31 VLT2000.05.N1.fits        
                                                                                 
 SOri38 &  904.12669251        &   49.0  6.0 & 2400  \\   %SOri38 VLT2000.01.N2.fits        
   ''   &  904.15630248        &   46.1  8.5 & 2400  \\   %SOri38 VLT2000.02.N2.fits        
   ''   &  904.19053588        &   49.0  6.4 & 2400  \\   %SOri38 VLT2000.03.N2.fits        
                                                                                 
 SOri39 &  905.13713833        &    8.5  3.2 & 2400  \\   %SOri39 VLT2000.02.N3.fits        
   ''   &  905.17120378        &    8.6  2.9 & 2400  \\   %SOri39 VLT2000.03.N3.fits        
   ''   &  905.20057526        &   10.1  3.0 & 2400  \\   %SOri39 VLT2000.04.N3.fits        
   ''   &  905.23010044        &   17.9  3.2 & 2400  \\   %SOri39 VLT2000.05.N3.fits        
   ''   &  905.25940527        &   11.5  3.9 & 2400  \\   %SOri39 VLT2000.06.N3.fits        
   ''   &  905.28869611        &    6.0  3.3 & 2400  \\   %SOri39 VLT2000.07.N3.fits        
   ''   &  905.31784490        &    9.8  5.0 & 2400  \\   %SOri39 VLT2000.08.N3.fits        
                                                                                 
 SOri42 &  904.05650249        &   88.3 12.5 & 2400  \\   %SOri50 VLT2000.00.N2.fits 
                                                                                 
 SOri44 &  903.23775122        &    3.3  1.2 & 4800  \\   %SOri44 VLT2000.05.N1.fits        
                                                                                 
 SOri45 &  903.23775122        &   26.4 15.4 & 4800  \\   %SOri45 VLT2000.05.N1.fits        
                                                                                 
 SOri49 &  906.11072522        &    4.2  3.7 & 4800  \\   %SOri47 VLT2000.ave.N4.fits
\hline
\end{tabular}
$\,$\\   %
\end{table*}

\setcounter{table}{2}
\begin{table*}
\caption[]{H$\alpha$ equivalent widths and observing dates for the  $\sigma$ Orionis cluster
brown dwarf candidates.  }
\begin{tabular}{llrl}
\hline
name    &     MJD-51000      & pW(H$\alpha$)& Exp.Time \\   %    
        &    (days)          &  (\AA)       & (seconds)\\   %
    (1) &      (2)           &  (3)         &    (4)                      \\   %
\hline      
SOri-J053826.1-024041 &903.05138618       &    3.3  1.2 & 2400 \\   % SOri47 VLT2000.00.N1.fits iac4-015
       ''             &906.03996572       &    6.1  2.2 & 2400 \\   % iac4-015
       ''             &906.07400201       &    3.7  2.1 & 2400 \\   % iac4-015
       ''             &906.11072522       &    7.8  2.9 & 2400 \\   % iac4-015
       ''             &906.14487566       &    1.9  0.7 & 2400 \\   % iac4-015
       ''             &906.17892987       &    1.3  0.7 & 2400 \\   % iac4-015

SOri-J053909.9-022814 &903.10135488 & $<$0.0   -- & 2400 \\   % so1-1-3199      
       ''             &903.14482793 & $<$0.0   -- & 2400 \\   % so1-1-3199      
       ''             &903.17441719 & $<$0.0   -- & 2400 \\   % so1-1-3199      

SOri-J053911.4-023333 &903.10135488 &    5.3  1.2 & 2400 \\   % so1-1-3113      
       ''             &903.14482793 &    4.9  0.8 & 2400 \\   % so1-1-3113      
       ''             &903.17441719 &    5.4  1.5 & 2400 \\   % so1-1-3113      
                                                    
SOri-J053912.8-022453 &904.12669251 &    2.5  0.7 & 2400 \\   % so1-1-3055      
       ''             &904.15630248 &    1.6  0.9 & 2400 \\   % so1-1-3055      
       ''             &904.19053588 &    2.8  1.3 & 2400 \\   % so1-1-3055      
                                                                                             
SOri-J053948.1-022914 &904.22764064 &    6.2  3.6 & 2400 \\   % iac5-041         
       ''             &904.26190299 &    4.8  3.5 & 2400 \\   % iac5-041        
       ''             &904.29136284 &    9.1  4.1 & 2400 \\   % iac5-041        
       ''             &904.32053636 &    6.5  4.3 & 2400 \\   % iac5-041        
                                                    
%%  SOri71            &904.05650249 &  704.7 74.0 & 2400 \\   % sok05-0586
\hline
\end{tabular}
$\,$\\   %
\end{table*}

\setcounter{table}{3}
\begin{table*}
\tiny
\caption[]{Spectral properties and classification of low mass stars and
 BDs of  $\sigma$ Orionis. We only list those members which have been observed 
spectroscopically.}
\begin{tabular}{l l cccc l}
\hline
 Name             &MK  &K band&H$\alpha$$^{*}$&Forbidden &Lithium & Type  \\
                  &    &excess&               &lines     &        &    \\
\hline

r053820-0234         &M4   &N/A & above & N  & Y  & CTT-  \\  %& 14.58  no IR data
SOri25               &M7.5 & N  & above &N/A &N/A & CTT-  \\  %& 17.04  
SOri38               &M7   & N  & above &N/A &N/A & CTT-  \\  %& 17.70  
SOri40               &M7   & N  & above &N/A &N/A & CTT-  \\  %& 17.93 
SOri66               &L3.5 & N  & above &N/A &N/A & CTT-  \\  %& 23.42  
SOri71$^\ddagger$    &L0   &N/A & above &N/A &N/A & CTT-  \\  %& 20.02 sok05-0586       
SOriJ053951.6-022248 &M5.5 & N  & above & N  & Y  & CTT-  \\  %& 14.59  
SOriJ054005.1-023052 &M5   & N  & above & N  & Y  & CTT-  \\  %& 15.90  

SOri12$^\dagger$     &M6   & Y  & below &N/A &N/A & CTT?  \\  %& 16.28 
SOri47 (UKIRT)       &L1.5 & Y  &  N/A  &N/A &N/A & CTT?  \\  %& 24.50 

SOri42               &M7.5 & Y? & above &N/A &N/A & CTT?? \\  %& 19.01 
SOriJ053912.8-022453 &M6   & Y? &  N/A  &N/A &N/A & CTT?? \\  %& 19.42 so1-1-3055       
SOriJ053948.1-022914 &M7   & Y? &  N/A  &N/A &N/A & CTT?? \\  %& 18.76 soIAC5-041       

4771-41              &K7   & Y  & above & Y  & Y  & CTT+  \\  %& 12.82  
r053833-0236         &M3.5 &N/A & below & Y  & Y  & CTT+  \\  %& 13.71  no IR data
r053840-0230         &M0   & Y  & below & Y  & Y  & CTT+  \\  %& 12.80  
r053849-0238         &M0.5 & N  & below & Y  & Y  & CTT+  \\  %& 12.88  
SOri45               &M8.5 & N  & above & Y  & Y  & CTT+  \\  %& 19.59  
SOriJ053827.4-023504 &M3.5 & N  & above & Y  & Y  & CTT+  \\  %& 14.50  
SOriJ053949.3-022346 &M4   & N  & above & Y  & Y  & CTT+  \\  %& 15.14  
SOriJ054001.8-022133 &M4   & Y  & above & Y  & Y  & CTT+  \\  %& 14.32  

SOri26               &M4.5 &N/A & below &N/A &N/A & NM?   \\  %& 17.05 
SOri49               &M7.5 &N/A & below &N/A &N/A & NM?   \\  %& 20.83 

4771-1038            &K8   & N  & below & N  & Y  & WTT   \\  %& 12.78  
4771-1051            &K7.5 &N/A & below & N  & Y  & WTT   \\  %& 12.33  
4771-1075            &K7   & N  & below & N  & Y  & WTT   \\  %& 12.66  
4771-1097            &K6   &N/A & below & N  & Y  & WTT   \\  %& 12.43  
4771-899             &K7   & N  & below & N  & Y  & WTT   \\  %& 12.08  
r053831-0235$^{**}$  &M0   & Y  & below & N  & Y  & WTT   \\  %& 13.52  high reddening
r053838-0236         &K8   & N  & below & N  & Y  & WTT   \\  %& 12.37  
r053907-0228         &M3   & N  & below & N  & Y  & WTT   \\  %& 14.33  
r053923-0233         &M2   & N  & below & N  & Y  & WTT   \\  %& 14.16  
SOri27               &M7   & N  & below & N  & Y  & WTT   \\  %& 17.07 
SOriJ053715.1-024202 &M4   & N  & below & N  & Y  & WTT   \\  %& 15.07  
SOriJ053820.1-023802 &M4   & N  & below & N  & Y  & WTT   \\  %& 14.41  
SOriJ053847.5-022711 &M5   & N  & below & N  & Y  & WTT   \\  %& 14.46  
SOriJ053914.5-022834 &M3.5 & N  & below & N  & Y  & WTT   \\  %& 14.75  
SOriJ053920.5-022737 &M2   & N  & below & N  & Y  & WTT   \\  %& 13.51  
SOriJ053958.1-022619 &M3   & N  & below & N  & Y  & WTT   \\  %& 14.19  

SOri11               &M6   & N  & below &N/A &N/A & WTT?  \\  %& 16.15   No Li measured
SOri13               &M5.5 & N  & below &N/A &N/A & WTT?  \\  %& 16.35   No Li measured
SOri15               &M5.5 & N  & below &N/A &N/A & WTT?  \\  %& 16.55   No Li measured
SOri17               &M6   & N  & below &N/A &N/A & WTT?  \\  %& 16.80   No Li measured
SOri20               &M5.5 & N  & below &N/A &N/A & WTT?  \\  %& 16.88   No Li measured
SOri22               &M6   &N/A & below &N/A &N/A & WTT?  \\  %& 16.98   No Li measured
SOri29               &M6.5 & N  & below &N/A &N/A & WTT?  \\  %& 17.18   No Li measured
SOri30               &M6   & N  & below &N/A &N/A & WTT?  \\  %& 17.25   No Li measured
SOri31               &M7   & N  & below &N/A &N/A & WTT?  \\  %& 17.29   No Li measured
SOri39               &M6.5 & N  & below &N/A &N/A & WTT?  \\  %& 17.82   No Li measured
SOri44               &M7   &N/A & below &N/A &N/A & WTT?  \\  %& 19.39   No Li measured
SOriJ053826.1-024041 &M8   & N  & below &N/A &N/A & WTT?  \\  %& 16.95   No Li measured soIAC4-015       
SOriJ053829.0-024847 &M6   & N  & below &N/A &N/A & WTT?  \\  %& 17.04   No Li measured so4-2-0731       
SOriJ053909.9-022814 &M5   & N  & below &N/A &N/A & WTT?  \\  %& 16.49   No Li measured
SOriJ053911.4-023333 &M5   & N  & below &N/A &N/A & WTT?  \\  %& 16.73   No Li measured
\hline
\end{tabular}
$\,$ \\
$^{*}$ W(H$\alpha$) above or below the criteria defined in the text 
(see Barrado y Navascu\'es et al. 2003).\\
$^{**}$ Large reddening.\\
$^\dagger$ ($K-L'$) excess from Jayawardhana et al. (2002)\\
$^\ddagger$ Asymmetry in H$\alpha$,  Barrado y Navascu\'es et al. (2002a)\\
CTT+      = Probable Classical TTauri star or analog, based on forbidden emission lines.\\
CTT-      = Possible Classical TTauri star or analog, based on H$\alpha$.\\
CTT?      = Possible Classical TTauri star or analog, based on the IR excesses.\\
CTT??     = Possible Classical TTauri star or analog? based on possible IR excesses.\\
WTT       = Weak-line TTauri star or analog based, on Lithium and H$\alpha$.\\
WTT?      = Possible Weak-line TTauri star or analog, based on H$\alpha$.\\
NM?       = Non-member?.\\
N/A       = Data no available.\\
\end{table*}